\documentclass{aa}  

\usepackage{graphicx}
\usepackage{txfonts}
\usepackage{lipsum}
\usepackage{subcaption}         
\usepackage{lscape}             
\usepackage{placeins}           
\usepackage{comment}
\usepackage{dsfont}
\usepackage{braket}

\begin{document}

   \title{Quantum-optimal coronagraphy with spatial mode sorting for direct exoplanet observations}

   \titlerunning{Quantum-optimal coronagraphy with spatial mode sorting}


   \author{Y. Xin\inst{1}\thanks{Corresponding author: xin@strw.leidenuniv.nl}
        \and S.Y. Haffert\inst{1}
        \and Y.J. Kim\inst{2}
        \and J. Lin\inst{3}
        }

   \institute{Sterrewacht Leiden, PO Box 9513, Niels Bohrweg 2, 2300 RA Leiden, The Netherlands
   \and Department of Physics \& Astronomy, 430 Portola Plaza, University of California, Los Angeles, USA
   \and NASA Ames Research Center, Moffett Boulevard, Moffett Field, USA
   }

   \date{Received September 30, 20XX}

 
  \abstract
   {Conventional coronagraphs struggle to reach the theoretical limit of exoplanet detection at close separations to the star, particularly when the telescope has a complex aperture or when the star is partially resolved. Coronagraphy or nulling using spatial mode-sorting can reach the theoretical limit, but the optimal solution has so far only been calculated for an idealized unresolved star, whose signal lies entirely in the piston mode of the telescope.}
   {This work aims to enable the calculation of optimal nulling modes for realistic observational scenarios as a function of the size of the star and planet parameters, with the goal of improving coronagraphic performance at $\sim \lambda/D$ working angles given partially resolved stars and complex telescope apertures.} 
   {We perform numerical calculations using tools from quantum information theory and explore the behavior of optimal mode-sorting measurements.}
   {The optimal measurement for measuring a planet parameter is calculable from the density matrix describing the state of the system. The spatial mode that maximizes the classical signal-to-noise ratio is approximately quantum optimal to leading order in the stellar leakage and the planet flux ratio. We present optimal modes for measuring planets with known signals, and we characterize the tradeoffs inherent to coronagraphs targeting more than one planet location. Example coronagraph designs are presented for three cases of scientific interest: 1) the optimal extension of the fiber nuller architecture for detecting and spectrally characterizing planets across an arbitrary field-of-view using high-resolution spectroscopy, 2) following up planets detected by the visible coronagraph of the Habitable Worlds Observatory at more challenging infrared wavelengths, and 3) detecting and localizing planets at close working angles with the Planetary Camera and Spectrograph on the Extremely Large Telescope.}
   {}


   \maketitle


\nolinenumbers

\section{Introduction}

Directly detecting the light of exoplanets is very challenging due the extreme flux ratios between the faint planets and their bright host stars, which range from on the order of $10^{-4}$ in the infrared for hot young planets to $10^{-10}$ in visible wavelengths for mature rocky planets \citep{bowler_imaging_2016,currie_direct_2023}. Additionally, the planets are spatially close to their host star, and large telescopes are required to resolve them as distinct point sources. However, directly measuring the light of Earth-like exoplanets is crucial for finding signs of life, as its spectral properties allow us to detect molecular biosignatures in the planet’s atmosphere \citep{national_academies_of_sciences_engineering_and_medicine_pathways_2021}. Two upcoming observatories capable of directly measuring Earth-like planets are the Extremely Large Telescope (ELT) and the Habitable Worlds Observatory (HWO). Both will use an instrument called a coronagraph to suppress the light of the star in the optical system before it reaches the detector \citep{kenworthy_high-contrast_2025}. However, no existing coronagraph design has yet been demonstrated in laboratory settings at the performance levels needed to detect exo-Earths. The most promising progress for coronagraphs with very deep stellar suppression has been at visible wavelengths, with high contrast demonstrations of coronagraphs that can access planets beyond 2-4 $\lambda/D$, but attenuate throughput of closer-in planets. HWO will need to characterize exoplanets up to $1.7 \mu$, wavelengths \citep{krissansen-totton_wavelength_2025}, and most of the target planets accessible by conventional coronagraph designs at visible wavelengths will appear around $1 \lambda/D$ in the near-infrared, where planet throughput would be heavily attenuated by those same coronagraph designs. Meanwhile, on the ELT, accessing mature rocky exoplanets with an instrument such as the Planetary Camera and Spectrograph (PCS) will also require $10^{-8}$ levels of stellar suppression at $1 \lambda/D$ \citep{kasper_pcs_2021}. Therefore, designing coronagraphs that can achieve deep stellar suppression while maintaining planet throughput at $1 \lambda/D$ is a key technology gap for both HWO and the ELT.

Conventional coronagraph architectures struggle to reach the required performance at these close-in separations, especially when partially resolved stars and complex telescope apertures (such as those of HWO and the ELT) are considered \citep{kenworthy_high-contrast_2025}. However, there has been recent interest in directly measuring exoplanets at the quantum limit \citep{deshler_quantum_2025}. Our work aims to expand the use of concepts and tools from quantum sensing to enable the design of coronagraphs that can achieve the theoretical performance limit set by fundamental physics. This work is applicable for observing planets at any separation, but the largest gains over existing coronagraph designs occur at the close-in separations that are most critical for the search for life in the coming years.

The most widely used coronagraphs to date typically filter out starlight through a series of pupil and focal plane masks, optionally with pupil-reshaping lenses in front to apodize the beam \citep{kenworthy_high-contrast_2025}. After the starlight is removed from the optical system, a focal plane containing mostly off-axis light is imaged onto a detector grid. The final data typically resembles that of an image of the astrophysical scene, discretized by the detector pixels. This is especially true far away from the star where the impact of the coronagraph on the point-spread-function (PSF) is minimal. However, there have long existed coronagraphs that do not form data resembling that of imagers, for example the Visible Nuller Coronagraph, and more recently, a series of coronagraph designs based on inducing patterns onto the optical beam with an complex transmission mask, then coupling it into fibers \citep{haguenauer_deep_2006,ruane_efficient_2018, por_single-mode_2020}. The fiber coupling operation is mathematically described as an overlap integral between the electric field $E$ at the input of fiber with the (complex conjugate of the) mode profile $\phi$ belonging to the fiber, with the measured intensity corresponding to the squared-amplitude of the integral: 

\begin{equation}
    I = \bigg|\iint_D \phi^*(\vec{\xi}) E(\vec{\xi}) d^2\xi \bigg|^2,
\end{equation}

where both $\phi$ and $E$ are each of unit-norm. This overlap integral can also be thought of as a projection of the electric field $E$ onto the spatial mode $\phi$. Similarly, an instrument that injects light into multiple fibers, or into a mode-sorting device such as a photonic lantern \citep{birks_photonic_2015}, can be thought of as projection of $E$ onto a set spatial modes $\phi_k$. Fiber-based coronagraphs are thus natural implementations of spatial mode demultiplexers (SPADEs) \citep{tsang_quantum_2016}, which has received interest in the quantum imaging community for its superresolution capabilities. However, there is nothing uniquely quantum about a SPADE relative to conventional coronagraphs. In fact, even imaging a focal plane onto a detector is a SPADE: one that projects the focal electric field onto the modes described by top-hat functions corresponding to each detector pixel.

Despite there being no fundamental physical difference between conventional coronagraphs, fiber-based coronagraphs, and the SPADEs used in superresolution work, quantum information theory provides powerful tools for the analysis, design, and characterization of coronagraphs \citep{huang_ultimate_2023, deshler_quantum_2025}. For an introduction to quantum sensing and the quantum-mechanical notation used in this paper, as well as discussions about the connection between quantum information theory and conventional coronagraphy metrics such as signal-to-noise ratio, we refer the reader to Appendices A-C. The rest of this work will assume familiarity with the concepts and notation presented therein.

Under the quantum sensing framework, non-orthogonal measurements (known as positive operator-valued measurements) are possible; however, optimal coronagraph design corresponds to the optimal projective-value measure --- the optimal SPADE \citep{tsang_quantum_2016, deshler_quantum_2025}. One physical implementation of a (non-pixel-based) SPADE that measures an arbitrary single spatial mode is coupling the input light into a structured waveguide that only accepts the desired mode. A SPADE measuring multiple modes at once can be realized using mode-sorting devices such as multi-plane light converters \citep{morizur_programmable_2010, labroille_efficient_2014}, photonic integrated circuits \citep{norris_first_2020, martinod_scalable_2021,sirbu_astropic_2024}, or photonic lanterns \citep{birks_photonic_2015}. Theoretically, arbitrary mode-sorting can also be realized with a bulk-optics combination of Mach-Zehnder interferometers, though this is typically complicated in practice and susceptible to path-length instabilities.

Given these recent advancements in the technologies that enable spatial mode demultiplexing, it becomes important to identify the optimal modes to sort. \citet{deshler_quantum_2025} showed that in the idealized scenario of an unresolved, infinitesimally small star, in the limit of extreme flux ratios $c\rightarrow0$, and in the limit of the separation between the planet and the star being much smaller than the diffraction limit, then the quantum optimal measurement is that of any signal orthogonal to the fundamental piston mode. \citet{deshler_quantum_2025} showed that in this idealized scenario, this optimal single-state projection saturates the quantum limit given by the quantum Chernoff bound (QCB) \citep{audenaert_discriminating_2007}, the maximum possible rate of exponential decay in the probability of error when hypothesis testing between the presence or absence of a planet.

In our work, we extend the formalism to investigate more realistic astronomical scenarios and enable the calculation of optimal projective measurements as a function of stellar radius and planet parameters. We focus our analysis to the practical design of mode-sorting coronagraphs, which can only realize repeated projective measurements of single-photon states. Although repeated projections on single-photon states are not guaranteed to saturate the QCB given finite-sized stars \citep{huang_ultimate_2023}, such measurements can achieve quantum optimal sensitivity in parameter estimation tasks \citep{helstrom_quantum_1969}, such as the measurement of the planet flux. We argue that in the context of coronagraph design, optimally measuring the planet flux is intuitively connected to the classical notion of signal-to-noise ratio, and is a more useful definition of optimality than minimizing the hypothesis testing error.

\section{Theory of optimal measurements} \label{sec:theory}
A primer on the relevant notation and selected concepts from quantum sensing can be found in App. \ref{app:qm_primer}. In this section, we apply known tools and results from quantum sensing to the coronagraphy problem. Following \citet{deshler_quantum_2025}, but using the slightly different notation of $\vec{\xi}$ to denote the on-sky spatial coordinate, we write the density matrix describing the state of the incoming photon in the presence of a planet at $\vec{\xi}_p$, with relative flux contribution $c$, as

\begin{equation} \label{eq:density_matrix}
\begin{split}
    \hat{\rho} &= (1-c)\ket{\psi(\vec{\xi}_s)}\bra{\psi(\vec{\xi}_s)}+c\ket{\psi(\vec{\xi}_p)}\bra{\psi(\vec{\xi}_p)} \\
    &= (1-c)\ket{\psi_s}\bra{\psi_s}+c\ket{\psi_p}\bra{\psi_p}.
\end{split}
\end{equation}

This density matrix represents the state of a photon that has arrived from the planet with probability $c$, and from the star with probability $1-c$. If there is no planet, then this density matrix is still valid with $c=0$. Note that to reduce the number of variables used across this analysis, we have used $c$ to denote both the planet flux ratio ($c=I_p/I_s$ where $I$ is used to denote intensity) and the relative flux contribution of the planet ($c=I_p/(I_s+I_p)$), which is only valid if $I_p\ll I_s$. The optimal measurement for state discrimination, as calculated in \citet{deshler_quantum_2025}, is that of the projector onto the space orthogonal to $\ket{\psi_s}$.

We now generalize the density matrix for finite-size stars or unknown planet positions as

\begin{equation} \label{eq:density_matrix}
    \hat{\rho} = \frac{(1-c)}{N_s}\sum_i\ket{\psi_{s_i}}\bra{\psi_{s_i}}+\frac{c}{N_p}\sum_j\ket{\psi_{p_j}}\bra{\psi_{p_j}}.
\end{equation}

For simplicity, we've written the expression for the density matrix as if the planet were equally probable anywhere in the FOV spanned by $\ket{\psi_{p_j}}$, though it can be modified to have non-uniform probabilities if desired. The problem of detecting the existence of a planet is problem of discriminating between the state with $c=0$ (denoted as $\hat{\rho}_0$) and that with non-zero $c$ (denoted as $\hat{\rho}_1$). The measurement that minimizes the error probability when hypothesis testing between $\hat{\rho}_0$ and $\hat{\rho}_1$ given the measurement of a single photon is known as the Helstrom measurement \citep{helstrom_quantum_1969}, and is the projection onto the positive and negative eigenspace of $\hat{\rho}_1-\hat{\rho}_0$. Given $N$ copies of the state (i.e. $N$ photons) and the problem of distinguishing between $\hat{\rho_0}^{\otimes N}$ and $\hat{\rho_1}^{\otimes N}$, the optimal solution is the projective measurement on the positive and negative eigenspace of $\hat{\rho_1}^{\otimes N}-\hat{\rho_0}^{\otimes N}$. However, this solution is impractical for coronagraphy as it not only depends on $N$, but requires storing the states of $N$ incoming photons before performing the measurement. What is conventionally thought of as a coronagraph is a linear operator $C$ that can only repeat the same measurement for each incoming photon. The optimal coronagraph for minimizing classification error is thus the one that maximizes the classical Chernoff bound \citep{chernoff_measure_1952}, given by

\begin{equation} \label{eq:classical_chernoff}
    \xi_{\mathrm{CB}} = -\log (\min_{0 \leq s \leq 1}  \sum_k p_0(k)^s p_1(k)^{1-s})
\end{equation}

where the sum over $k$ is a sum over the measurement bases, and $p_0$ and $p_1$ the probability of measuring $k$ given $\hat{\rho}_0$ or $\hat{\rho}_1$ respectively. See App. \ref{app:chernoff_deriv} for a preliminary discussion of how this measurement would be calculated.

We pursue an alternative approach in this work, which is to maximize the sensitivity of the measurement to the parameter $c$. The optimal measurement for estimating $c$ is given by the symmetric logarithmic derivative (SLD) with respect to $c$, denoted as $\hat{L}_c$ and given by

\begin{equation}
    \partial_c \hat{\rho}(c) = \frac{1}{2}(\hat{L}_c \hat{\rho}(c) + \hat{\rho}(c)\hat{L}_c).
\end{equation}

The SLD, or any operator that commutes with it, is the optimal measurement that maximizes the classical Fisher information (CFI), which encodes the sensitivity of the measurement to the parameter $c$. The supremum of the CFI given all possible measurements is the quantum Fisher information (QFI), a quantity intrinsic to the state itself given by $F_Q(c) = \mathrm{Tr}(\hat{\rho}(c)\hat{L}^2_c)$. In the basis that diagonalizes the density matrix given by $\hat{\rho} = \sum_m \zeta_m \ket{\zeta_m} \bra{\zeta_m}$, the SLD can be calculated as

\begin{equation} \label{eq:sld_closed_form}
    \hat{L}_c = 2 \sum_{m,n} \frac{\braket{\zeta_n|\partial_c\hat{\rho_1}|\zeta_m}}{\zeta_m+\zeta_n} \ket{\zeta_n}\bra{\zeta_m}.
\end{equation}

As explored in App. \ref{app:quantum_to_classical}, the statistics of this measurement are the most closely related to the statistics encapsulated by the common classical expression for S/N in the presence of Poisson noise. Unlike the Helstrom measurement, this measurement is also optimal for estimating $c$ given $\hat{\rho}^{\otimes N}$ since the QFI scales linearly with the number of copies of the state. Therefore, although achieving the QFI for measuring $c$ is not strictly optimal for minimizing classification error for hypothesis testing, it is overall a more versatile notion of quantum optimality for coronagraphy.

When the position of the planet is completely unknown, as is the case with detection surveys, measurements optimized for constraining $c$ do not provide good constraints on what the planet location actually is. The calculations are nevertheless useful in specific instances, for example if the planet's location is already known (to some level of certainty) and we merely wish to maximize our ability to measure its brightness in the presence of photon noise. This scenario is explored for the case of the Habitable Worlds Observatory in Section \ref{sec:hwo}. Additionally, as explored in Section \ref{sec:fiber_nulling}, this formalism provides the optimal extension to fiber nulling \citep{haguenauer_deep_2006, ruane_efficient_2018}, where ports suppress starlight but couple planet light from some FOV to a spectrograph, where it is analyzed using high-resolution spectroscopy. In this observation mode, the molecular lines from species in the planet's atmosphere (which are not expected to exist in the star's atmosphere) are used to detect the planet \citep{wang_observing_2017}. These molecular features can be used to study the planet even if its location and absolute brightness remain unconstrained \citep{echeverri_vortex_2024}.

In general, there are performance tradeoffs when defining a finite field-of-view of a coronagraph, and the more uncertain the planet's location is, the less optimal the resulting solution is for any one particular planet. Coronagraphs that can also constrain the properties of the planet are even more complicated, with additional tradeoffs in how well different potential planet states can be distinguished from one another. These tradeoffs are explored in more detail in Section \ref{sec:fov_tradeoffs}.

\section{Coronagraphs for a single planet} \label{sec:single_planet_sld}

In this section, we numerically calculate the quantum optimal, maximum information measurement, given by the SLD. We model the star as a uniform disk of finite extent. Using Eq. \ref{eq:density_matrix}, from $\hat{\rho_1} = (1-c)\hat{\rho}_0+c\ket{\psi_p}\bra{\psi_p}$, we have $\partial_c \hat{\rho_1} = \ket{\psi_p}\bra{\psi_p} - \hat{\rho}_0$. To numerically compute $\hat{L}_c$, we first calculate the eigenvectors of $\hat{\rho_1}$ (which depend on $c$), then build the components of $\hat{L}_c$ in that basis according to Eq. \ref{eq:sld_closed_form}. Then $\hat{L}_c$ can be transformed back to either the pupil plane or the focal plane to obtain the desired measurement mode.

We model a system in which the star has radius $R_s=0.2$ and the planet is at a location of $\xi_{x_p} = 1.0$. We've chosen this configuration primarily because it is a challenging coronagraphy problem to showcase the power of mode-sorting relative to conventional coronagraph architectures, which is most pronounced for close-in planets with resolved stars. And, as shown in Section \ref{sec:science_examples} and discussed further in App. \ref{app:apertures}, the framework easily extends to arbitrary telescope apertures, circumventing the difficulties typically encountered when designing coronagraphs given central obstructions, support spiders, and segment gaps.

\begin{figure*}[t]
\begin{center}
    SLD modes for finite star ($R_s=0.2$) and planet at $\xi_{x_p}=1.0$ with $c=10^{-7}$\\
    \includegraphics[scale=0.33]{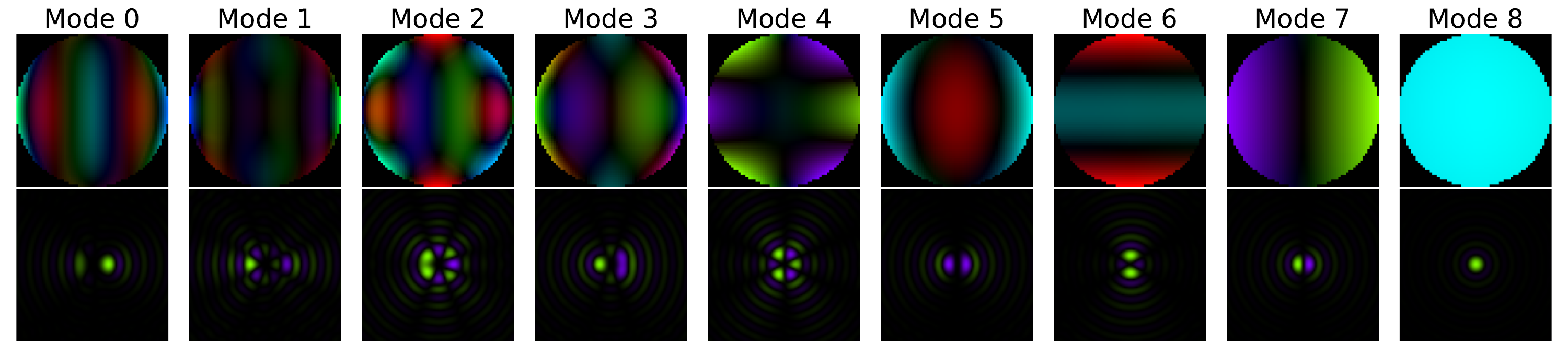}
	\caption{\label{fig:finite_star_sld} The eigenmodes of the SLD operator for a system with $R_s=0.2$, $\xi_{x_p}=1.0$, and $c=10^{-7}$, ordered from left to right by decreasing eigenvalue. We have omitted the axes label to conserve space; the pupil plane modes (top) are plotted from -0.5 to 0.5 in both axes, while the corresponding modes in the focal plane (bottom) are plotted from -10 to 10 in both axes. We see that the dominant Mode 0, which provides the most information about $c$, is very similar to the classically optimized mode for the same system shown in Fig. \ref{fig:finite_star_sols}. However, computing the SLD automatically gives us the remaining auxiliary modes in which we can measure additional information about $c$, as well as their relative informativeness, while jointly optimizing multiple modes using classical optimization techniques can be quite complicated.}
\end{center}
\end{figure*}

We next calculate the SLD operator for the same system with $c=10^{-7}$. This is numerically tricky as the nonzero eigenvalues of $\hat{\rho}_1$ span a large dynamic range (from $0.95$ to $10^{-37}$ for this example), but regularizing it by removing small eigenvalues can slightly change the resulting eigenvectors of the SLD. App. \ref{app:numerical_disc} contains a more detailed discussion about the numerical limits of this calculation. For all of the SLDs computed in this work, we achieve sensible results by building $\hat{\rho}$ using the first 1000 eigenmodes of the telescope, calculating the eigendecomposition of $\hat{\rho}$, then calculating the SLD using the modes of $\hat{\rho}$ with eigenvalues whose real part is greater than 0. The SLD we calculate then has many nonzero eigenvalues itself, but most of these are modes of just numerical noise. We thus filter for the eigenmodes of the SLD that have reasonable planet throughput, chosen to be above $\eta_p>10^{-3}$. While for some scenarios, there exist modes with $\eta_p<10^{-3}$ that are not purely numerical noise, we can numerically verify that these modes contribute a negligible amount to the overall QFI, so neglecting them does not impact the overall sensitivity. From a practical standpoint, modes with $\eta_p<10^{-3}$ are almost certainly not worth measuring given realistic levels of detector noise and finite detector dynamic range, especially since their contribution to the QFI would be infinitesmal. We show the resulting SLD modes in Fig. \ref{fig:finite_star_sld} and report the $\eta_{s_k}$ and $\eta_{p_k}$ corresponding to each port in Table \ref{tab:finite_star_sld_quants}.

The QFI is the CFI corresponding to the SLD, i.e. $F_Q = \sum_k (\partial_c p_k)^2/p_k = \sum_k (\eta_s-\eta_p)^2/(c\eta_p + (1-c)\eta_s)$. We also report in Table \ref{tab:finite_star_sld_quants} the individual QFI contribution $(\partial_c p_k)^2/p_k$ of each port, which is approximately the S/N ratio for $c$ of that port.

We see that the dominant mode of the SLD, i.e. the mode with the highest individual QFI contribution, is very similar to the classically optimized mode for the same system shown in the top right of Fig. \ref{fig:finite_star_sols}. The dominant mode only contributes about $80\%$ of the QFI in this case, since we have used a relatively large star with a close-in planet, as well as a finite $c$. We find that the relative contribution of the dominant mode increases as $c$ is decreased, contributing $88\%$ of the QFI if $c=10^{-12}$, keeping the other system parameters fixed. We also find that the relative contribution of the dominant mode increases as $R_s$ is decreased, contributing $99.5\%$ of the QFI if $R_s=0.02$, with $c=10^{-7}$ and the other parameters fixed --- or if the planet position is farther from the star, contributing $99.8\%$ of the QFI if the planet is moved to $2.0 \lambda/D$ with $R_s=0.2$, $c=10^{-7}$ and the other parameters fixed. In general, as either $R_s$ or $c$ approach 0, or as the planet grows further separated from the star, the quantum optimal solution becomes progressively better-approximated by the measurement of a single mode. This behavior agrees with analytical predictions, discussed further in App. \ref{app:quantum_to_classical}.

However, when the solution is not effectively singular, the SLD is a more powerful approach than conventional optimization techniques, since the SLD automatically gives us all the modes in which we can measure additional information about $c$, as well as their relative contribution to the QFI. If we can make a device that can sort any number of modes, the SLD tells us how many modes we need to achieve optimality. If we are restricted to the number of modes we can measure, then the relative QFI contributions tell us which ones are the best to keep. Although it is possible to conceive of the existence of these other modes using classical coronagraphy intuition, choosing the right number of modes and calculating them using classical optimization techniques would be significantly more complicated, requiring optimization over a manifold of orthonormal $\phi_k$ without prior knowledge of how many modes are needed. In certain scenarios, however, the SLD becomes so singular that the numerical calculation breaks down. This phenomenon is discussed in more detail in App. \ref{app:numerical_disc}. In this regime, we know the quantum optimal measurement is effectively a single mode, and the solution for the mode shape can be calculated using classical optimization techniques, such as the method described in App. \ref{app:classical_opt}.

On a qualitative note, when examining both the classically optimized modes and the SLDs, we find that for faint planets, it is more optimal to null the star at the expense of planet throughput. As the planet becomes brighter, retaining the planet light becomes more optimal. Unfortunately, the solutions do not converge to the same mode for the range of $c$ relevant for modern exoplanet science, so a real instrument would have to be optimized for a particular $c$ or for the average performance across a range of $c$, or else be designed to be reconfigurable for different values of $c$. For ground-based astronomy, the instrument will be limited by atmospheric residuals, so the instrument only has to be optimized for the range of $c$ that is both of scientific priority and accessible given the wavefront residuals.

\begin{table}[ht!]
\caption{Useful characteristic quantities for the SLD modes presented in Fig. \ref{fig:finite_star_sld}. The quantity $\eta_{p_k}$ is the planet coupling, $\eta_{s_k}$ the stellar coupling, and $(\partial_cp_k(c))^2/p_k(c)$, the contribution of the mode to the total QFI. This modal basis is optimal measurement for the specific system where $R_s=0.2$, $\xi_{x_p}=1.0$, and $c=10^{-7}$. The relevant values for the mode solution obtained using classical optimization techniques (see Fig. \ref{fig:finite_star_sols}), labeled cl, are also presented for comparison. The classically-optimized solution has the highest per-mode QFI, but the SLD provides higher total information.}

\label{tab:finite_star_sld_quants}
\centering
\begin{tabular}{c c c c}
\hline\hline
Mode ($k$) & $\eta_p(\xi_{p_x}=1.0)$ & $\eta_s$ & $(\partial_cp_k(c))^2/p_k(c)$\\
\hline
   cl & 0.033 & $1.46 \times 10^{-9}$ & $2.3 \times 10^{5}$ \\
   0 & 0.029 & $8.8\times10^{-10}$ & $2.2\times10^{5}$ \\    
   1 & 0.010 &  $8.2 \times10^{-9}$ & $1.2\times10^{4}$ \\
   2 & 0.025 &  $3.5 \times10^{-8}$ & $1.7\times10^{4}$ \\
   3 & 0.074 &  $2.3 \times10^{-7}$ & $2.3\times10^{4}$ \\
   4 & 0.062 &  $2.1 \times10^{-6}$ & $1.9\times10^{3}$ \\
   5 & 0.340 &  $1.3 \times10^{-4}$ & $9.0\times10^{2}$  \\
   6 & 0.064 &  $2.4 \times10^{-4}$ & $1.7\times10^{1}$ \\
   7 & 0.363&  $2.3 \times10^{-2}$ & $5.0\times10^{0}$ \\
   8 & 0.031&  $9.5 \times10^{-1}$ & $8.9\times10^{-1}$ \\
\hline                                  
\end{tabular}
\end{table}

\section{Theoretical limits of coronagraphy} \label{sec:coronagraph_limits}

\begin{figure*}[t]
\begin{center}
    a \includegraphics[scale=0.5]{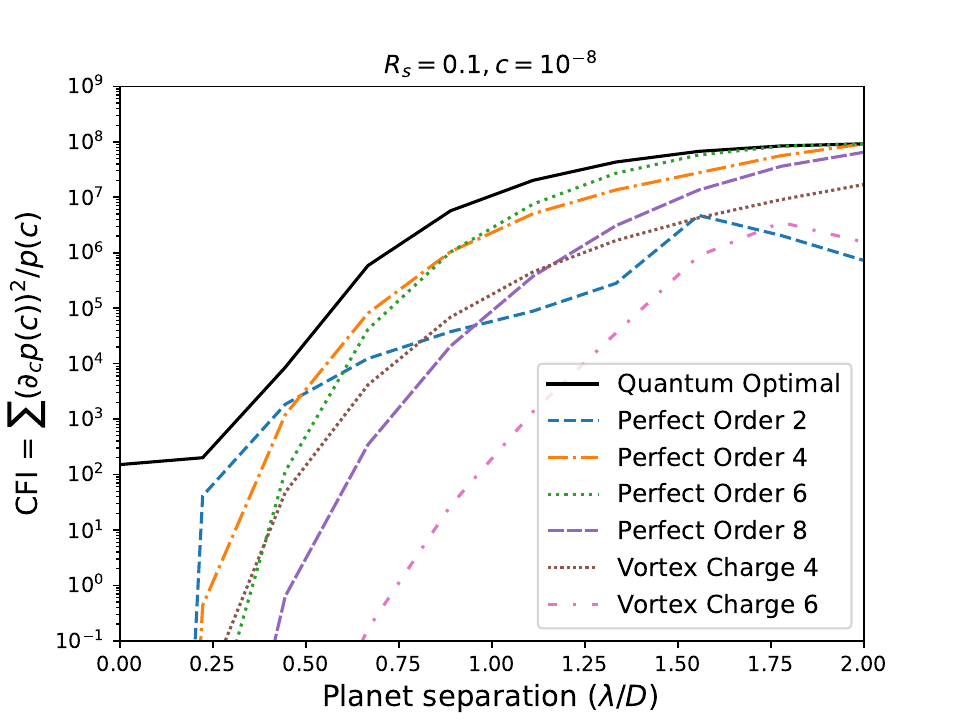}
    b \includegraphics[scale=0.5]{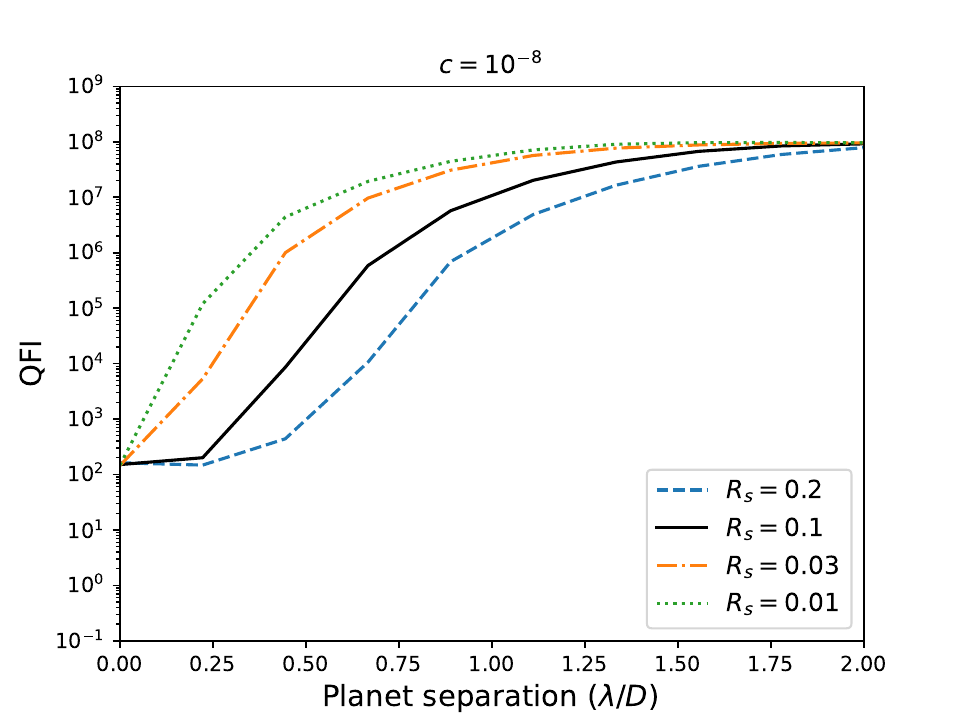}
	\caption{\label{fig:cfi_comparison} a) The QFI corresponding to the SLDs as a function of the planet separation, with $R_s=0.1$ and $c=10^{-8}$. The CFI of `perfect' coronagraphs that fully remove the lowest order modes up to (but excluding) orders 2, 4, 6, and 8 are also plotted, as well as the CFI of charge 4 and 6 vortex coronagraphs. The SLD achieves the highest sensitivity to $c$. Here, the calculations for the perfect and vortex coronagraphs assume that the light not removed by the coronagraph is projected onto the planet mode. Projecting the light onto a pixel basis (as is often assumed when referring to perfect coronagraphs) would result in even lower CFI. b) The QFI curves for a range of astrophysically relevant $R_s$ (with $c$ fixed at $10^{-8}$) plotted on the same scale as part (a). As the star becomes smaller, coronagraphs can better measure the $c$ of a planet at a given position. Beyond separations of $2 \lambda/D$, however, the difference is not very noticeable.}
\end{center}
\end{figure*}

The theoretical limit of coronagraphy (for sensitivity to the planet's relative intenstiy $c$) is given by the QFI, which is intrinsic to the state of the astrophysical system as viewed by the telescope aperture. In 
Fig. \ref{fig:cfi_comparison}, we show some example theoretical limits with a circular aperture as a function of the stellar radius, the relative flux of the planet, and the planet position. We also compare the QFI with the CFI of `perfect' coronagraphs that fully remove the lowest order modes up to (but excluding) orders 2, 4, 6, and 8, as well as charge 4 and charge 6 vortex coronagraphs \citep{mawet_annular_2005}. From Eq. \ref{eq:cfi_general} and Eq. \ref{eq:p_phi}, the CFI of any coronagraph with more than one output can be expressed in terms of the $\eta_{s_k}$ and $\eta_{p_k}$ of its outputs:

\begin{equation} \label{eq:cfi_etas}
    F_c(c) = \sum_k \frac{(\eta_{p_k}-\eta_{s_k})^2}{(1-c)\eta_{s_k}+c \eta_{p_k}}.
\end{equation}

We note that metrics such as summed throughput and averaged contrast (or null-depth) on their own are not useful or informative for mode-sorting coronagraphs, for which the range of $\eta_s$ and $\eta_p$ across the modes can span orders of magnitude. The CFI, which can be calculated from the value of $c$ and the $\eta_{s_k}$ and $\eta_{p_k}$ of the coronagraphic outputs (the pixels in the case of conventional imaging-type coronagraphs), provides a useful, generalized metric to rigorously compare coronagraphs of different architectures. Figure \ref{fig:cfi_comparison} shows that the QFI is indeed higher than the CFIs of the perfect-by-order coronagraphs and vortex coronagraphs. Additionally, our calculations for the perfect and vortex coronagraphs assume that the light not removed by the coronagraph is projected onto the planet mode. Projecting the light onto a pixel basis (as is often assumed when referring to perfect coronagraphs) would result in even lower CFI.

Unlike with the CFI of perfect-by-order coronagraphs, the QFI does not go to 0 when the planet position goes to the origin. This is because the presence of a planet at the origin will appear as an excess in the piston mode relative to the uniform disk of the star, such that its relative flux can still be measured.

\section{Measurement incompatibility with a finite field-of-view} \label{sec:fov_tradeoffs}

We next extend our analysis to coronagraphs with a finite field-of-view (FOV), which is generally desirable as we would like a single instrument capable of detecting planets at multiple potential locations and of constraining the planet's location and brightness. However, while it is possible to find an optimal measurement for measuring a planet of unknown location within some FOV, there is usually no single optimal solution that achieves simultaneous maximal S/N for measuring all possible states over the entire FOV. Additionally, even optimizing for one measurement across some defined FOV will usually result in suboptimal performance when considering any other FOV (including any subset of the defined FOV). The exception is when the FOV is chosen such that a very particular set of conditions are satisfied. This is a result of measurement incompatibility when estimating multiple parameters, e.g. when the optimal measurement $\hat{L}_{w_1}$for estimating parameter $w_1$ and the optimal measurement $\hat{L}_{w_2}$ for estimating parameter $w_2$ don't commute, and therefore are not simultaneously realizable \citep{ragy_compatibility_2016}.

To explore how measurement incompatibility manifests in coronagraphy, we can examine a system where the planet can be in two different states, $\psi_{p_1}$ with probability $c_1$ or $\psi_{p_2}$ with probability $c_2$, written in density matrix form as 

\begin{equation}
    \hat{\rho} = (1-c_1-c_2)\hat{\rho}_0+c_1\ket{\psi_{p_1}}\bra{\psi_{p_1}}+c_2\ket{\psi_{p_2}}\bra{\psi_{p_2}}.
\end{equation}

Optimizing measurements for estimating $c_1$ and $c_2$ can be thought of as designing a coronagraph with a FOV defined by the two spatial samples $\ket{\psi_{p_1}}$ and $\ket{\psi_{p_2}}$. For any combination of $c_1$, $c_2$, $\ket{\psi_{p_1}}$, and $\ket{\psi_{p_2}}$, we can calculate the SLDs $\hat{L}_{c_1}$ and $\hat{L}_{c_2}$. We can only simultaneously optimally measure $c_1$ and $c_2$ if, for commutator $[\hat{L}_{c_1},\hat{L}_{c_2}] = \hat{L}_{c_1}\hat{L}_{c_2}-\hat{L}_{c_2}\hat{L}_{c_1}$, we have $\mathrm{Tr}(\hat{\rho}(c_1,c_2)[\hat{L}_{c_1},\hat{L}_{c_2}]) = 0$. To characterize the level of incompatibility for measuring planets in two potential locations, we can calculate the quantumness parameter \citep{carollo_quantumness_2019, mihailescu_critical_2025}, defined as

\begin{equation}
    R = ||\frac{1}{2}\mathrm{Tr}(\hat{\rho}(c_1,c_2)[\hat{L}_{c_1},\hat{L}_{c_2}]) \hat{F}_{Q}^{-1}||_{\infty}.
\end{equation}

The $||\cdot||_\infty$ is the spectral norm, or the eigenvalue with the largest magnitude, and $\hat{F}_Q$ is the quantum Fisher information matrix \citep{helstrom_quantum_1969}, which can be calculated using the element-wise formula $\hat{F}_{{Q}_{j j'}} = \frac{1}{2}\mathrm{Tr}(\hat{\rho}\{\hat{L}_{j}\hat{L}_{j'}\})$, where the brackets indicate an anticommutator $\{A,B\} = AB+BA$.

\begin{figure}[ht!]
\begin{center}
    \includegraphics[scale=0.5]{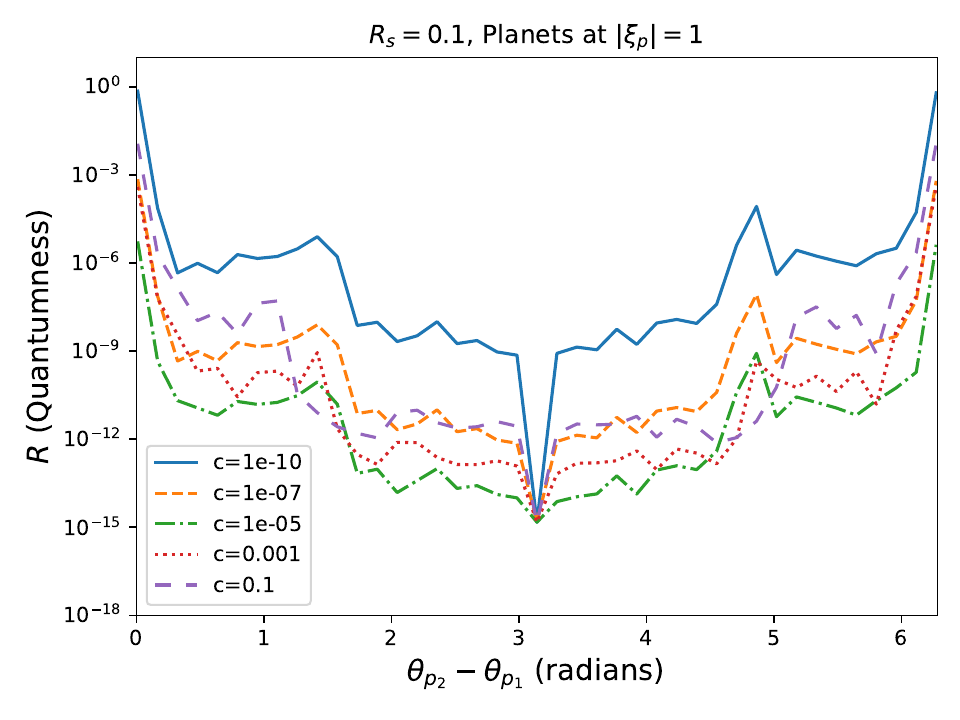}
	\caption{\label{fig:measurement_incompatibility} The quantumness parameter $R$, describing the level of measurement incompatibility between two potential planet signals. We hold one planet fixed at $|\xi_p|=1$ and $\theta_{p_1}=0$. The second planet also has a separation of $|\xi_p|=1$, and we calculate the incompatibility as a function of $\theta_{p_2}$. Because of various numerical limitations, we do not interpret our calculations of $R$ as quantitative measures of measurement incompatibility, but rather as a qualitative exploration of how the incompatibility between two potential planets varies with their brightness and the separation between them.}
\end{center}
\end{figure}

The value of $R$ can be between 0 and 1, with 0 corresponding to full compatibility and 1 corresponding to maximum incompatibility. We attempt to calculate some example values of $R$, holding $\ket{\psi_{p_1}}$ fixed at $|\xi_p|=1$ (with $\theta_{p_1}=0$). We vary the position angle corresponding to $\ket{\psi_{p_2}}$ but keep the separation the same. Unfortunately, the calculation of $R$ is extremely numerically fragile. Both $\hat{L}_{c_1}$ and $\hat{L}_{c_2}$ are near singular, and the subtraction of their pairwise multiples is very noisy. Additionally, $\hat{\rho}$ is extremely ill-conditioned, and the trace over its full rank is also very noisy. After various careful numerical considerations, including using a spectral formula for $\mathrm{Tr}(\hat{\rho}(c_1,c_2)[\hat{L}_{c_1},\hat{L}_{c_2}])$ adapted from \citet{ragy_compatibility_2016}, and restricting the trace to the effective rank \citep{roy_effective_2007} of $\hat{\rho}$, we obtain the values of $R(\hat{\rho}(c_1,c_2))$ shown in Fig. \ref{fig:measurement_incompatibility}.

Despite careful numerical treatment, there are remaining issues with the computed values of $R$. For a symmetric pupil, the value of $R$ is expected to be symmetric about $\theta_{p_2}-\theta_{p_1}=\pi$ (or zero), but we observe deviations from a fully symmetric curve. Given that the shape of the observed asymmetry is similar across multiple values of $c$, it may arise from a slight asymmetry in the input planet signals caused by the finite numerical precision of the values of $\theta_{p_2}$. We also note that the complex128 data type used for these calculations has a decimal precision of about 15, which results in the numerical floor at $R\sim10^{-15}$.

Because of these numerical limitations, we do not interpret our calculations of $R$ as quantitative measures of measurement incompatibility, but rather as a qualitative exploration of how the incompatibility between two potential planets varies with their brightness and mutual separation. The overall shape of $R$, rather than the exact values, is what is useful for informing coronagraph design, i.e. the balance between FOV coverage with the S/N achievable for any given location. In App. \ref{app:numerical_disc}, we discuss the physical factors behind the difficulty of these computations as well as potential alternative approaches in limiting cases.

Although we have used quantum sensing terminology to quantify the measurement incompatibility between different points in the FOV, this incompatibility has physically intuitive classical interpretations. For example, in the hypothetical limit of $c\gg1$, the optimal measurement of an off-axis source would be projection of the incoming electric field onto the electric field of that source. Meanwhile, projecting the electric field onto a detector pixel grid in the focal plane (i.e. forming an image) achieves near-uniform sensitivity over the spatial coordinates, but is suboptimal for all of them. In the regime where the optimal measurement reduces to that of a single mode, an intuitive interpretation of measurement incompatibility is the inability to stack fibers with these modes arbitrarily close to each other. When such fibers are placed at locations where their modes would be non-orthogonal, the modes will actually interfere with one another such that the measured modes become orthogonal, and therefore no longer as optimal. A related effect is also seen in the calculation of optimal mode-shapes for single-mode integral-field spectroscopy, in which the optimal modes for densely sampled fields are uniform modes (e.g. detector pixels), and while the optimal modes for sparsely sampled fields resembles that of typical single-mode fibers (SMFs) \citep{haffert_fundamental_2021}. In the coronagraphy context, however, the optimal mode shape is not only dependent on the planet throughput but also on its cancellation of the star, and as shown in work with fiber-injection units \citep{mawet_observing_2017}, sparsely spaced modes can allow for additional stellar suppression relative to the dense and uniform pixel modes.

In some cases, it is practical to choose a design that is suboptimal if it has other desirable properties. At larger working angles, the conventional coronagraph architecture that projects a filtered focal plane electric field onto a detector may still be the most practical kind of coronagraph, as it can implemented with conventional focal plane science detectors and produces data that is sufficiently related to the astrophysical scene to be interpreted by eye. It is, however, suboptimal from an S/N perspective for measuring the flux of any particular planet. In general, there is no single optimal coronagraph that achieves maximum S/N for all possible planets, and the coronagraph designer has to manage the inherent tradeoffs of their desired FOV and impose the relevant properties of the instrument depending on their use case.

\section{Examples} \label{sec:science_examples}
In this section, we show a few example cases for which the analysis machinery of mode-sorting coronagraphy is extremely powerful, whether the resulting measurements are actually implemented using mode-sorting devices or approximated using more conventional architectures. We do not attempt to find here the best designs for each science case, each of which would be a significant project on its own that involves careful consideration of the target list, the detectors available and their properties, the observatory or mission concept of operations, the wavefront errors involved, the wavefront control systems, and many other complicated factors. Rather, we present toy examples motivated by the overall goals and needs of each science case, showing how the analysis framework can be used to study fundamental tradeoffs in coronagraphy. We show in practice how the optimal coronagraph changes based on how the problem is defined, emphasizing that the specific demands placed on the coronagraphic system will impact the degree to which \textit{any} coronagraph is capable of achieving them. For pedagogical reasons, we first examine the case of fiber nulling on the ELT, showing how as the FOV targeted by the nuller is reduced, the performance of the nuller within the FOV improves. We then examine the limiting case where the planet location is known, as might be the case when planets detected using the visible wavelength coronagraph of HWO need follow-up spectral characterization at more challenging infrared wavelengths. Lastly, we return to the ELT to examine the case of multiparameter estimation that is necessary for both detecting and localizing exoplanets, such as a potential survey to be conducted by the PCS instrument.

\subsection{Optimal fiber nulling for arbitrary FOVs} \label{sec:fiber_nulling}

The Palomar Fiber Nuller \citep{haguenauer_deep_2006} uses two subpupils of the telescope aperture, phase-shifted by $\pi$ with respect to each other such that their on-axis beams, when injected into a SMF, will destructively interfere. Meanwhile, off-axis beams will receive a different relative phase shift, allowing some planet light to couple into the fiber. The fiber nuller (FN) architecture has since expanded to include other phase masks \citep{ruane_efficient_2018}. Although FN provides a measurement in which the planet flux is degenerate with its position, at high spectral resolutions, planets can be detected and characterized through the spectral features molecules in their atmospheres.

The physical implementation for measuring one arbitrary mode can be approximately achieved using a single pupil-plane mask (e.g. a complex apodizer) plus a focusing optic that injects the beam into a conventional SMF. This architecture is practically desirable if the planet light is to be fed through the fiber into a high-resolution spectrograph, such as in the case of many existing and planned instruments \citep{delorme_keck_2021, konopacky_development_2023, vigan_first_2024, kasper_pcs_2021}. There also now exist fiber-based coronagraphs with multiple outputs. The single-mode complex mode refinement (SCAR) coronagraph \citep{por_single-mode_2020, haffert_single-mode_2020}, which consists of a bundle of fibers with a phase mask to null the fields that couple into the fibers, is one of the first fiber-based instruments that allowed for the measurement of multiple modes simultaneously, with planet localization capability. The photonic lantern nuller (PLN) \citep{xin_efficient_2022}, which uses a mode-selective photonic lantern to cancel on-axis starlight, is another expansion of the FN architecture that also measures multiple modes simultaneously and partially localizes the planet.

Most previous work on FNs has usually first presented a phase mask designed using interferometric principles and subsequently characterized the performance and working FOV of the design. The SCAR coronagraph, meanwhile, is designed using numerical optimization techniques that maximize planet throughput while constraining contrast. However, as observed in Section \ref{sec:single_planet_sld}, constraining the level of stellar suppression a priori does not lead to information-optimal coronagraphs. Additionally, the architecture of the SCAR coronagraph also constrains the space of measurement modes that can be realized.

In this work, we explore how calculating the SLD allows for the design of the FN that is optimal from an information perspective, regardless of physical architecture, for any defined FOV. In particular, we examine the case of a FN on the ELT, which can potentially be implemented as an observation mode of the PCS instrument. The stellar radius is set to $R_s=0.1$, though this can be adjusted depending on the average stellar radius of the scientific target list. The target flux ratio is set to $c=10^{-8}$, corresponding to the type of rocky planets that PCS aims to observe. We first analyze the results for a FOV defined as a uniformly sampled full circle of radius $|\xi_p|=1$, corresponding to the situation in which we have no information about the position angle of the planet. We also analyze the results for a FOV with $|\xi_p|=1$, but extending only from $\theta_p=0$ to $\theta_p=\pi/2$. This is relevant if the approximate quadrant that the planet lies in is known.

\begin{figure*}[t]
\begin{center}
    a) \includegraphics[scale=0.3]{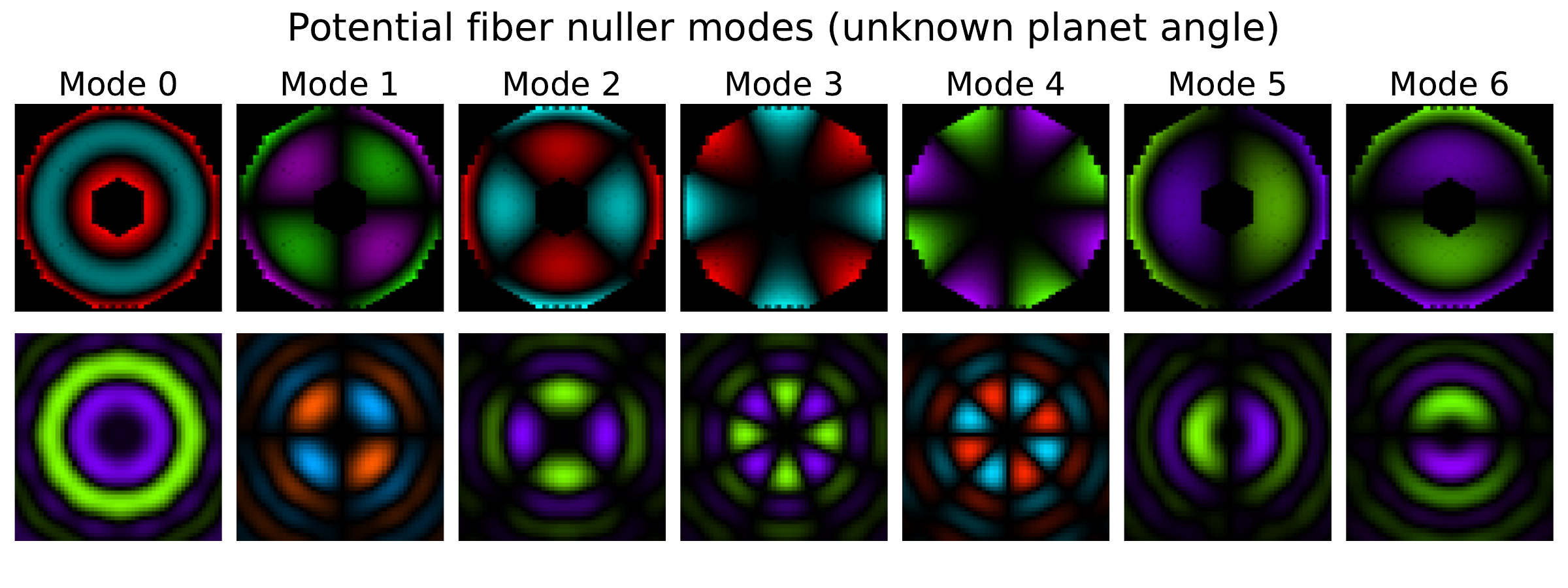}
    b) \includegraphics[scale=0.58]{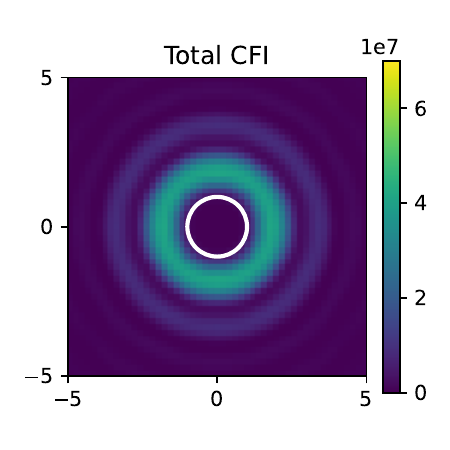}
	\caption{\label{fig:fn_circle} a) The first 7 eigenmodes of the SLD operator for a system with $R_s=0.1$, a full-circle FOV with $|\xi_p|=1.0$, and $c=10^{-8}$. The remaining eigenmodes contribute relatively less information so are not shown. The modes are calculated assuming the aperture of the ELT, and are ordered from left to right by decreasing eigenvalue. We have again omitted the axes label to conserve space; the pupil plane modes (top) are plotted from -0.5 to 0.5 in both axes, while the corresponding modes in the focal plane (bottom) are plotted from -5 to 5 in both axes. b) The total CFI over all SLD modes, as a function of the spatial separation. The QFI of the measurement problem corresponds to the CFI on the design FOV defined by the $|\xi_p|=1.0$ circle (denoted by the white line). The coronagraph is only quantum-optimal on this circle, but it also provides high information at farther separations.}
\end{center}
\end{figure*}

The density matrix for each FOV is generated according to Eq. \ref{eq:density_matrix}, where the $\psi_{p_j}$ are chosen to sample the desired FOV. Then, the SLD that optimizes sensitivity to the parameter $c$ is calculated. We present the first 7 SLD modes which have $\eta_p > 10^{-3}$ for the full-circle FOV in Fig. \ref{fig:fn_circle}a. The remaining eigenmodes contribute relatively less information and are not shown. Although the design FOV is defined by $|\xi_p|=1$, the SLD modes have higher planet throughput elsewhere. In terms of classical coronagraphy metrics, Modes 0 achieves $\eta_s=3.68 \times 10^{-13}$ and a maximum $\eta_{p}$ on the $|\xi_p|=1$ ring of 0.0032, with planet throughput over all space reaching a maximum of $\eta_p=0.044$ at $|\xi_{p}|= 3.3$. Similarly to what is observed in Table \ref{tab:finite_star_sld_quants}, the smaller eigenvalue modes tend to have less deep $\eta_s$ and higher $\eta_p$. However, rather than listing $\eta_s$ and the $\eta_p$ of various locations for all of the ports, it is more concise now to report the performance of this coronagraph using a spatial map of the CFI, calculated at each point using Eq. \ref{eq:cfi_etas}. This spatial map, which shows the sensitivity to $c$ provided by this coronagraph as a function of the planet location, is shown in Fig. \ref{fig:fn_circle}b.


\begin{figure}[ht!]
\begin{center}
    \includegraphics[scale=0.6]{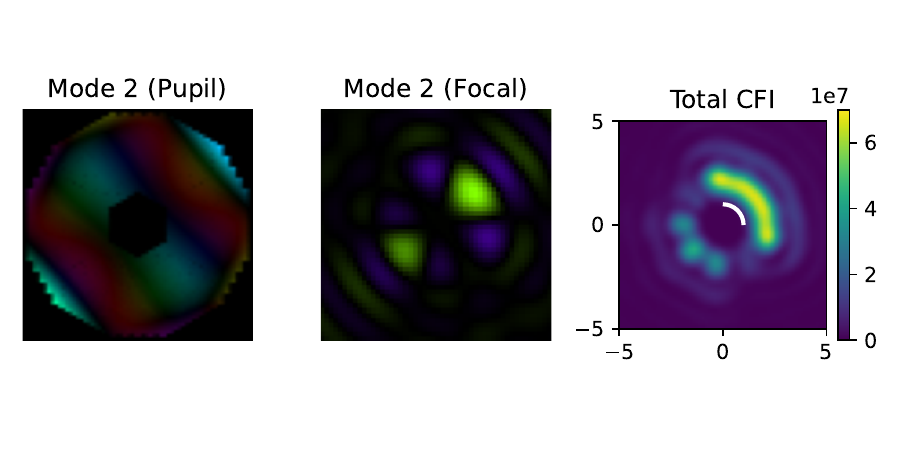}
	\caption{\label{fig:fn_quadrant} SLD calculations with $R_s=0.1$, $c=10^{-8}$, and $|\xi_{p}|=1$ as in Fig. \ref{fig:fn_circle}, but for a smaller FOV over a quadrant from $\theta_p = [0,\pi/2)$. Mode 2 has the highest individual QFI contribution, with $\eta_s=2.4\times10^{-11}$, a maximum $\eta_p$ on the $|\xi_p|=1.0$ arc of 0.027, and a maximum $\eta_p$ over all space of 0.45.  We have omitted the axes label to conserve space; the pupil plane mode (left) is plotted from -0.5 to 0.5 in both axes, while the corresponding mode in the focal plane (middle) is plotted from -5 to 5 in both axes. The other modes of the SLD, with lower individual QFI contributions, are not shown. The spatial map of CFI across all modes (right) is plotted on the same scale as in Fig. \ref{fig:fn_circle}, showing that a reduction in the targeted FOV (indicated with the white line) increases the achievable sensitivity within that FOV.}
\end{center}
\end{figure}

In Fig. \ref{fig:fn_quadrant}, we show example results from restricting the FOV to be the top right quadrant of the $|\xi_p|=1.0$ ring. In this example, Mode 2 has the highest individual QFI contribution. In terms of classical coronagraphy metrics, it achieves $\eta_s=2.4\times10^{-11}$, a maximum $\eta_p$ on the $|\xi_p|=1.0$ arc of 0.027, and a maximum $\eta_p$ over all space of 0.45. We omit showing the other modes for the sake of conciseness, but as before, we also plot in Fig. \ref{fig:fn_quadrant} the spatial map of total CFI across all modes.

Both examples achieve higher CFI at spatial separations farther than the design FOV than at the design FOV itself, where the measurement is quantum optimal. This illustrates the fundamental information constraint at small separations due to the finite-sized star. We also see that reducing the target FOV (by reducing the uncertainty on the position of the planet in the formulation of the density matrix), we are able to achieve coronagraphic measurements with deeper $\eta_s$, higher $\eta_p$ in the defined FOV, and thus higher QFI at the defined FOV --- as well as higher CFI at larger spatial separations where the throughput peaks. Overall, this calculation of the SLD provides the optimal basis in which to measure the value of $c$, given some uncertainty in the planet position. However, based on the symmetries of the modes, we can see that this measurement does not provide much ability to distinguish among the different potential planet locations.

\subsection{Infrared follow-up coronagraph for the Habitable Worlds Observatory} \label{sec:hwo}

The Habitable Worlds Observatory mission aims to detect and characterize Earth-like planets around sun-like stars, searching their atmospheres for potential signs of life. To properly constrain atmospheric abundances, HWO needs to obtain spectral information at ultraviolet, visible, and infrared wavelengths, and a long-wavelength cutoff of $1.7\mu$m has been recommended based on retrieval studies \citep{krissansen-totton_wavelength_2025}. To date, most of the efforts in coronagraphy for HWO has focused on the visible. The potential targets used for yield studies are usually chosen such that their habitable zones are observable with leading visible coronagraph designs, often designed for separations of $2.5-10$ $\lambda/D$ with planet throughput significantly attenuated below $2.5 \lambda/D$ \citep{belikov_coronagraph_2024}. However, these habitable zones appear much closer to the star at infrared wavelengths, with a majority at separations of $1-2.5$ $\lambda/D$ \citep{stark_paths_2024}.

Attenuating the light of partially resolved stars while maintaining planet throughput at close-in separations --- especially with a segmented and obstructed aperture --- has traditionally been a difficult coronagraphy problem \citep{belikov_theoretical_2021}. However, if the visible wavelength coronagraph is able to localize the planets, then the infrared coronagraph does not necessarily have to cover a large FOV. It only has to measure the spectra of the planets at infrared wavelengths, with as high S/N as possible. As we have shown, reducing the FOV improves the coronagraphic performance within the FOV, so reducing the FOV to the minimum necessary to achieve the mission goals will make high S/N spectra as attainable as possible.

In this section, we show the quantum optimal measurement for a hypothetical HWO planet of known location given a potential telescope aperture. The shape of HWO's primary mirror has not yet been decided, so we use the aperture of the James Webb Space Telescope (JWST), which is segmented and obstructed by the secondary mirror and its support structures. This aperture has similar characteristics to the USORT aperture studied by the coronagraph design survey, but it can be more accessibly generated by the community using the make\_jwst\_aperture function from hcipy. We assume a stellar radius of $0.03$ $\lambda/D$ (at the recommended infrared long wavelength cutoff of $\sim 1.7 \mu$m) and place the design planet at $\xi_{x_p} = 1$ $\lambda/D$, corresponding to the closest-in and most difficult to observe habitable zones.

\begin{figure*}[t]
\begin{center}
    a) \includegraphics[scale=0.4]{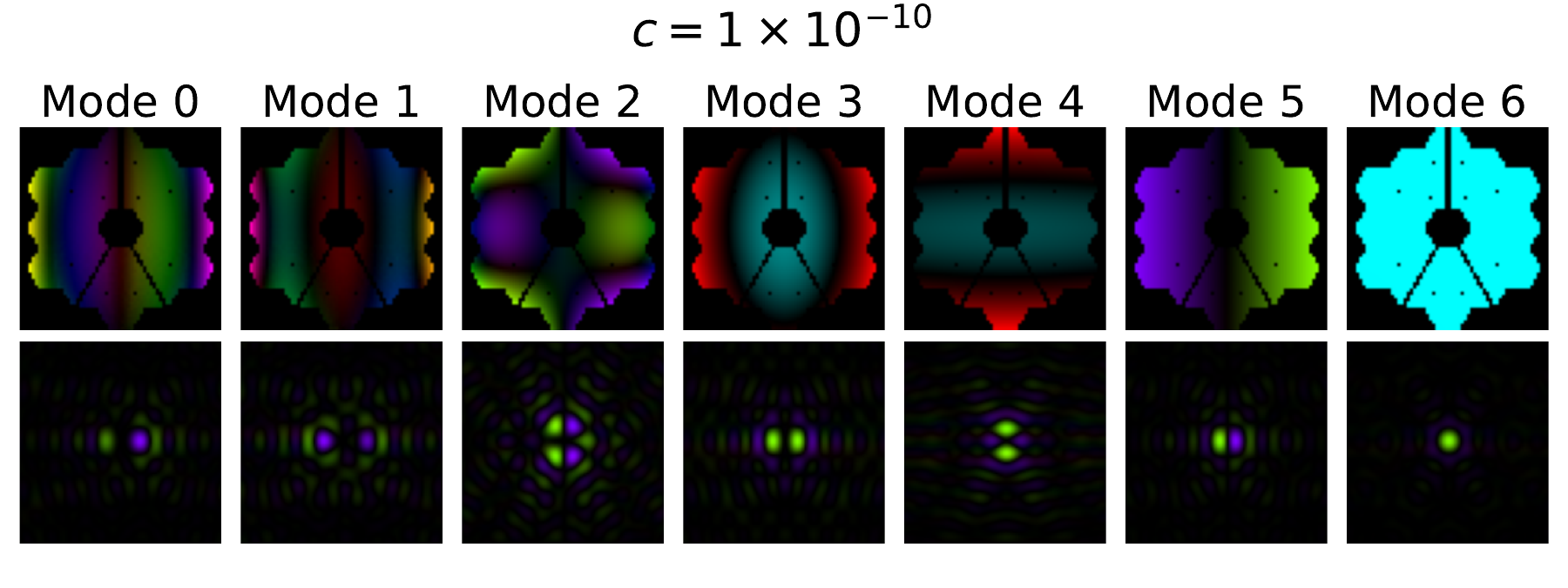}
    b) \includegraphics[scale=0.58]{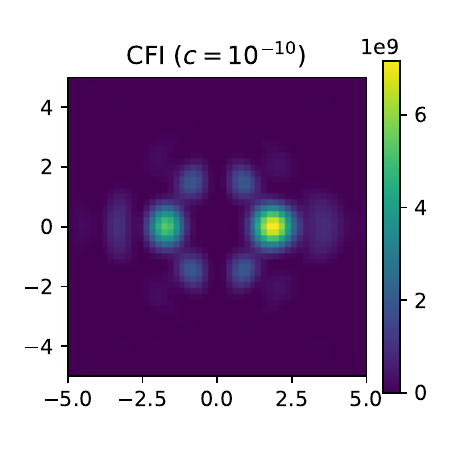}\\
    c) \includegraphics[scale=0.4]{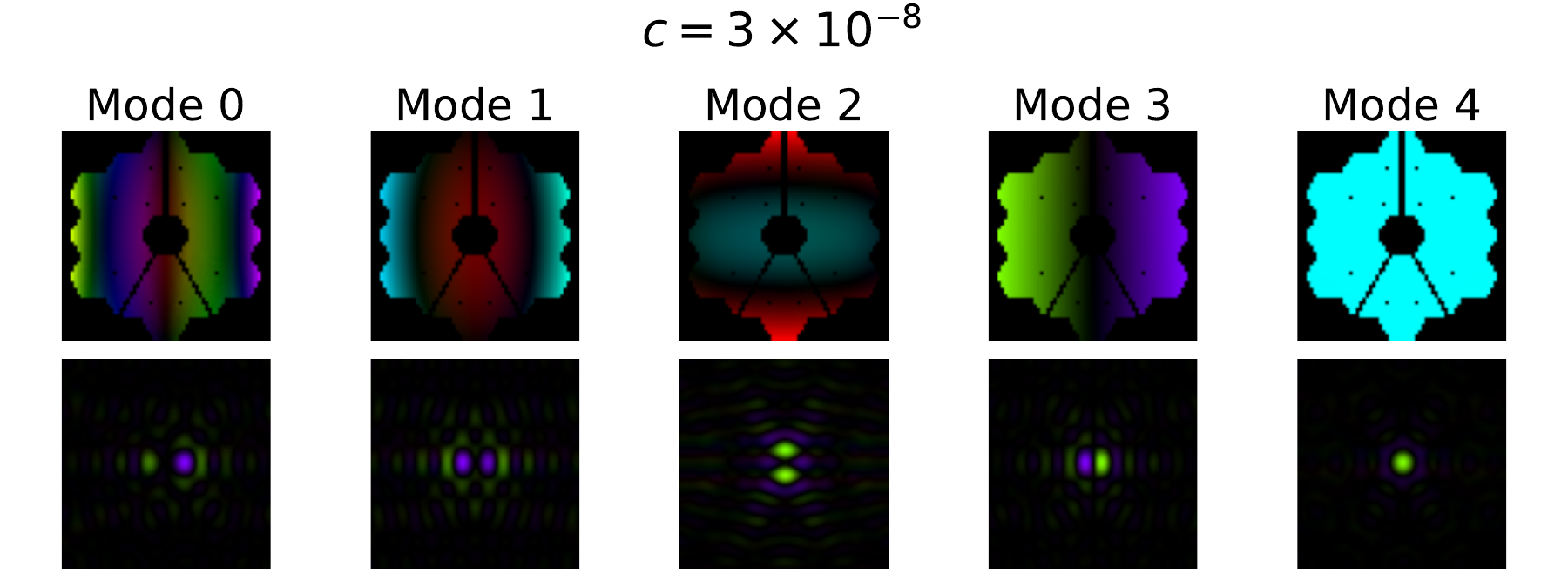}
    d) \includegraphics[scale=0.62]{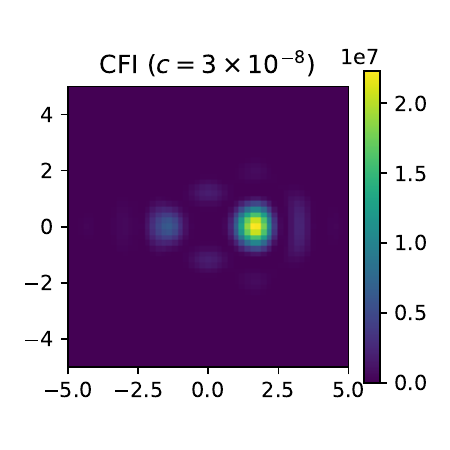}
	\caption{\label{fig:hwo_slds} a) The eigenmodes of the SLD operator for a system with $R_s=0.03$, a planet at $\xi_{p_x}=1.0$, and $c=10^{-10}$, given a JWST-like aperture. The modes are ordered from left to right by decreasing eigenvalue. We have again omitted the axes label to conserve space, but the pupil plane modes (top) are plotted from -0.5 to 0.5 in both axes, while the corresponding modes in the focal plane (bottom) are plotted from -5 to 5 in both axes. The dominant mode contributes $99.8\%$ of the QFI and would make for a near-optimal coronagraph on its own. b) The CFI given by the SLD in part (a), as a function of the spatial location of the planet. c) The eigenmodes of the SLD operator for the same system, except with the design $c$ increased to $c=3\times10^{-8}$. This has the effect of trading some of the depth of $\eta_s$ for more planet throughput. d) The CFI given by the SLD in part (c), as a function of the spatial location of the planet.}
\end{center}
\end{figure*}

We first show the resulting SLD for the astrophysically motivated value of $c=10^{-10}$ in Fig. \ref{fig:hwo_slds}a. This measurement involves a dominant mode with $\eta_s=2.1\times10^{-12}$ and $\eta_p(\xi_{p_x}=1) = 0.12$. In reality, we expect exozodical light to limit the achievable sensitivity at IR wavelengths to well above $10^{-12}$ levels \citep{ertel_hosts_2020, ertel_review_2025}, a limitation that, unlike wavefront errors, cannot be calibrated away. While exozodiacal light can be rigorously accounted for in the problem statement, here we choose to increase the design value of $c$ to $3\times10^{-8}$. This effectively trades some of the depth of $\eta_s$ for more planet throughput. The resulting SLD is shown in Fig. \ref{fig:hwo_slds}b, with the dominant mode having $\eta_s=1.1\times10^{-9}$ and $\eta_p(\xi_{p_x}=1)=0.193$. For the $c=10^{-10}$ SLD, the dominant mode contributes $99.8\%$ of the QFI and so would make for a near optimal coronagraph on its own. For the $c=3\times10^{-8}$ SLD, the dominant mode contributes $81.2\%$ of the QFI. Measuring Mode 1 in addition to Mode 0 would bring their combined QFI contribution up to $99.7\%$, at the cost of increased implementation complexity.

\begin{figure*}[t]
\begin{center}
    \includegraphics[scale=0.65]{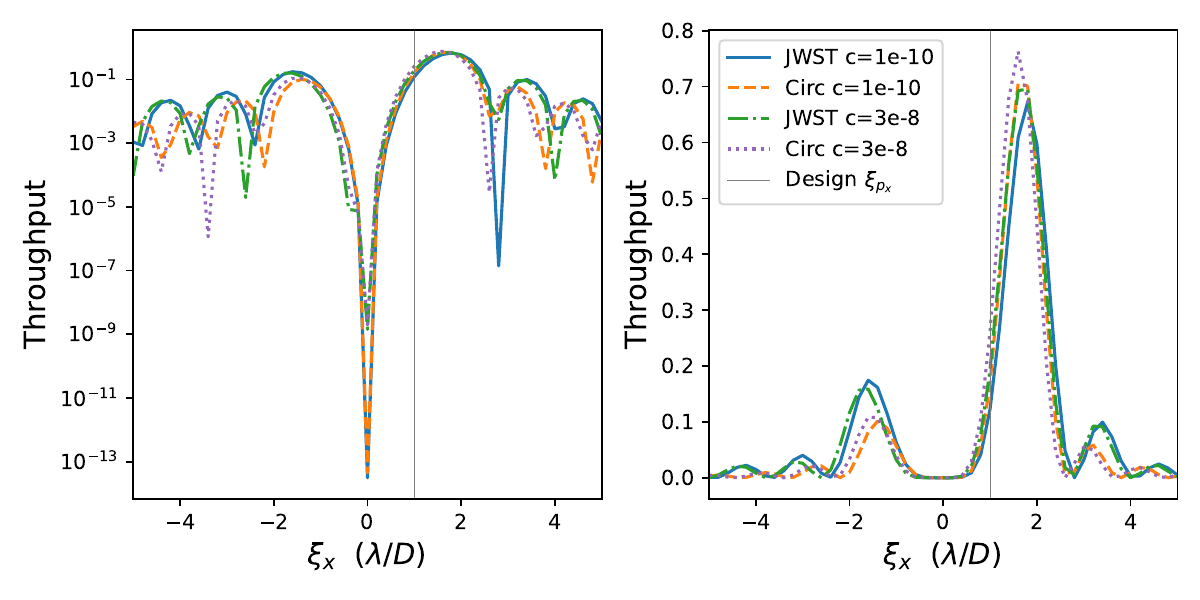}
	\caption{\label{fig:hwo_cross_sections} The cross-sectional throughputs in semilogy scale (left) and linear scale (right) of the dominant modes of the SLDs presented in Fig. \ref{fig:hwo_slds}, optimized for a JWST-like aperture with design values of $c=10^{-10}$ and $c=3\times10^{-8}$. Plotted for comparison are the throughputs of the equivalent modes optimized for a circular pupil that circumscribes the JWST aperture. The stellar radius is assumed to be $R_s=0.03 \lambda/D$ at HWO's long-wavelength cutoff of $1.7\mu$m. The SLD is optimized for $\xi_{p_x}=1.0 \lambda/D$ (indicated by the vertical gray line), though the resulting mode would also be quite good up until about $2.5 \lambda/D$.}
\end{center}
\end{figure*}

We plot the values of $\eta_s$ and $\eta_p$ (along the X-axis) for the dominant modes in Fig. \ref{fig:hwo_cross_sections}, noting that even though we calculated the modes that are quantum optimal for a planet at $\xi_{p_x}=1.0$, the planet throughput is actually higher at farther separations. This mode would thus make for a quite good, if somewhat suboptimal, coronagraph for planet separations up until about $2.5 \lambda/D$ 
We also present the corresponding throughput curves for a circular pupil that circumscribes the JWST aperture, confirming previous theoretical predictions \citep{belikov_theoretical_2021} that the aperture shape does not significantly impact the fundamental sensitivity limits. See also App. \ref{app:apertures} for a comparison of the fundamental limits across several different apertures.

We have assumed that the relative angle of the coronagraph to the sky can be properly aligned, either through internal optics (like a rotating mask mount) or through telescope rolls, such that the planet falls at the right angle to be maximally coupled. However, if the relative angle cannot be arbitrarily chosen, then it is possible to have either an adaptable coronagraph (e.g. one with multiple phase masks such that one can always be chosen to observe a planet at any given position angle) or a more complicated mode-sorting coronagraph that can measure planet light from multiple locations simultaneously. In general, measurements of multiple planet signals cannot be simultaneously optimal, and even sampling that satisfies $[\hat{L}_{c_j},\hat{L}_{c_{j'}}]=0$ is not expected to result in the best scientific yield. Rigorously choosing $\ket{\psi_{p_j}}$ and their respective weights to optimize for scientific yield is a topic left for future work. For a preliminary solution involving the use of the Pretty Good Measurement (PGM) \citep{montanaro_distinguishability_2007} for quantum state discrimination, see Section \ref{sec:pcs_surveyor}.

Overall, targeting one or a few position angles will result in higher coronagraphic performance at those positions than a coronagraph optimized for a uniformly-sampled $360^{\circ}$ FOV. However, this approach assumes that the location of the planet at the time of the infrared-wavelength observation can be accurately predicted from the orbital constraints obtained from visible-light imaging, which still needs to be tested with operations-level studies that include orbit-fitting. Another consideration is that if there are two or more habitable-zone planets around the same star, using limited samples that do not align well with the relative positions of the planets could potentially cost more telescope time to characterize the whole system. Further analysis of how frequently this situation would arise, given planet multiplicity statistics, would be necessary to inform the kind of IR coronagraph that is necessary to accomplish mission goals.

\subsection{Survey coronagraph for the Planetary Camera and Spectrograph} \label{sec:pcs_surveyor}

We now turn again to PCS but with a different problem statement. Rather than just coupling light from some FOV into a spectrograph (while nulling the star), we would like to be able to detect unknown planets as well as constrain their location. We now write our density matrix as 

\begin{equation} \label{eq:multiparam_rho}
    \hat{\rho} = \frac{(1-\sum_jc_j)}{N_s}\sum_i\ket{\psi_{s_i}}\bra{\psi_{s_i}}+\sum_jc_j\ket{\psi_{p_j}}\bra{\psi_{p_j}}.
\end{equation}

Note that compared to Eq. \ref{eq:density_matrix} used in the analysis of fiber nulling in Section \ref{sec:fiber_nulling}, the values of $c_j$ here are not scaled by $N_p$. The difference is that in Eq. \ref{eq:density_matrix}, we are trying to measure the relative flux $c$ of one planet, which has probability $1/N_p$ of existing in each of the locations defined by $\ket{\psi_{p_j}}$. In Eq. \ref{eq:multiparam_rho}, we are trying to distinguish amongst multiple potential planets, each with relative flux $c_j$.

As mentioned previously, it is possible to sample the FOV such that $[\hat{L}_{c_j},\hat{L}_{c_{j'}}]=0$ is satisfied for all pairwise combinations of points. Choosing such samples is not necessarily the best strategy, as suboptimally measuring more points might result in overall better scientific yield. Future work on coronagraph design for PCS will optimize the FOV samples and their relative weights in conjunction with yield studies. For now, we present a simple mode-sorting coronagraph design based on a geometrically motivated choice of sampling.

\begin{figure}[ht!]
\begin{center}
    \includegraphics[scale=0.55]{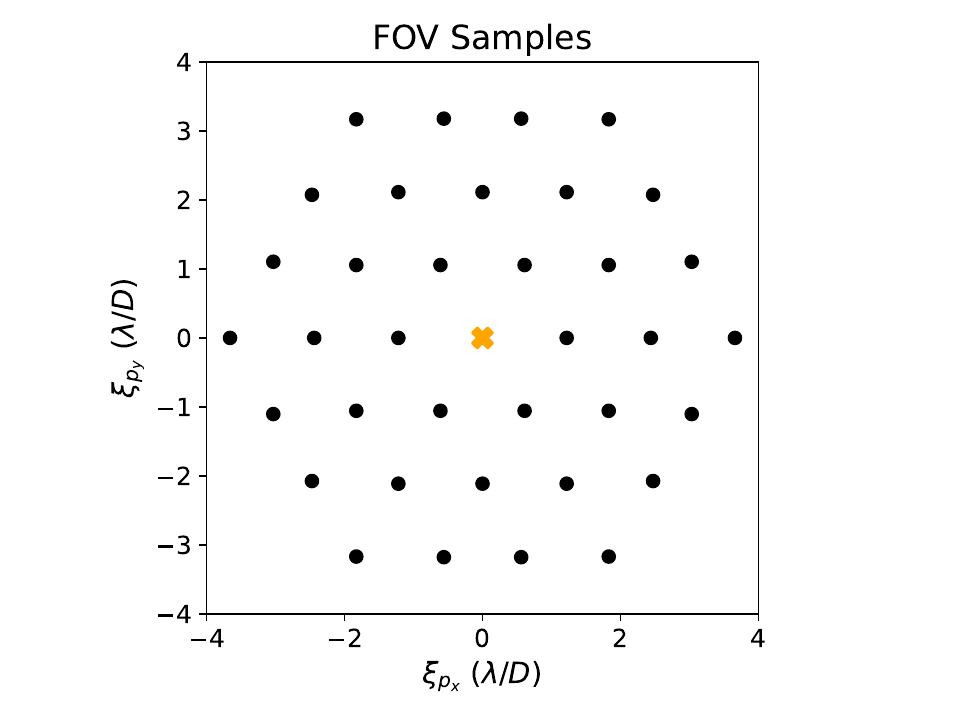}
	\caption{\label{fig:pcs_surveyor_fov} The FOV samples used in the calculation of the PGM modes shown in Fig. \ref{fig:pcs_surveyor_modes}. The samples are spaced out by $1.22 \lambda/D$ in a hexagonal pattern, such that for an ideal Airy PSF, the planet signal at each sample would be orthogonal to each of its nearest neighbors.}
\end{center}
\end{figure}

When the star is unresolved, there exist planet signals that are naturally fully orthogonal to the stellar signal. These points correspond to the dark rings of the Airy disk: the points where the stellar electric field in the focal plane crosses zero. For a circular pupil, the first ring occurs at the value of the Rayleigh criterion given by $1.22 \,\lambda/D$. Because planet signals share the same shape as the stellar signal but translated in space in the focal plane, the planet signals are orthogonal to each other if separated by $1.22 \, \lambda/D$. We can arrange our FOV samples in the hexagonal pattern shown in Fig. \ref{fig:pcs_surveyor_fov} such that the samples are all orthogonal to their nearest neighbor (for a circular pupil), or nearly so (for a close-to-circular pupil such as the ELT). Since the second and higher nulls of the Airy pattern are not at exact multiples of $1.22 \, \lambda/D$, these samples are not all orthogonal to each other, nor are they all orthogonal to the star. However, this choice of sampling removes the largest contributions to measurement incompatibility between the samples in the FOV, which arises from their adjacent samples. This layout is relatively straightforward to implement, even if in reality the telescope is not perfectly circular and the star will be partially resolved.

To calculate measurement modes for this FOV sampling, we borrow another tool from quantum information, the Pretty Good Measurement. For a state that is a sum over mixed states, $\hat{\rho} = \sum_l p_l \hat{\rho}_l$, one can attempt to distinguish between the states $\hat{\rho}_l$ using the PGM, which is given by

\begin{equation} \label{eq:pgm}
    \hat{M}_l = \hat{\rho}^{-1/2}(p_l\hat{\rho_l})\hat{\rho}^{-1/2}.
\end{equation}

In our case, $p_0\hat{\rho}_0$ is the state of the star, and $p_{(l\neq0)}\hat{\rho}_{(l\neq0)}=c_{(j=l)}\ket{\psi_{p_j}}\bra{\psi_{p_j}}$. Since $\hat{\rho}_{(l\neq0)}$ is rank 1, Eq. \ref{eq:pgm} results in a set of projectors where the measurement of each planet location corresponds to the measurement of one mode. This PGM is not the optimal measurement for state distinction, but it is much more straightforward to calculate and has been shown to have a success probability for state discrimination of at least the square of the optimal probability, with the relative success probability improving as the target states become more orthogonal to each other. Additionally, even though measuring $c_j$ is not explicitly the goal of the PGM, measuring $c_j$ is related to state distinguishability in the limit of small $c$, so we obtain modes that also have high S/N for measuring $c_j$.

For numerical calculations of the PGM, we use the innermost three layers of hexagonal FOV samples, as shown in Fig. \ref{fig:pcs_surveyor_fov}. These 36 samples cover a region that is typically hard to access using conventional coronagraphs. More layers of samples can be included if a wider FOV is desired. However, under this sampling framework, there would be progressively more samples per layer at larger separations. Given the complexity of sorting additional modes with one device, it might be more practical to use a conventional coronagraph beyond $4 \lambda/D$ if its performance is sufficient for achieving the science goals.

\begin{figure*}[t]
\begin{center}
    \includegraphics[scale=0.6]{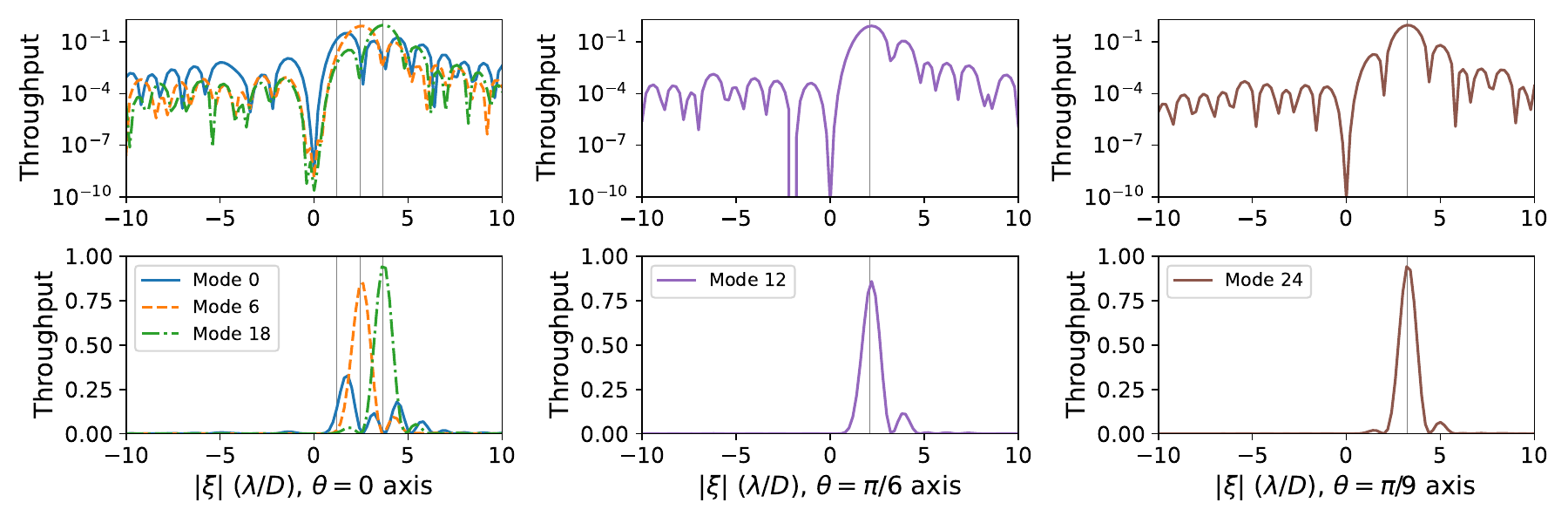}
	\caption{\label{fig:pcs_surveyor_xsec} The cross-sectional throughputs in semilogy scale (top) and linear scale (bottom) of the unique PGM modes, along the axis that intersects the corresponding FOV sample. The other PGM modes are rotations of the ones shown, so have the same radial throughput profiles.}
\end{center}
\end{figure*}

The 36 modes onto which we should project, calculated using Eq. \ref{eq:pgm}, are shown in Fig. \ref{fig:pcs_surveyor_modes} in App. \ref{app:pcs_pgm_modes}. The throughput cross-sections along the axis intersecting the FOV samples are shown in Fig. \ref{fig:pcs_surveyor_xsec} for the unique modes, numbered 0, 6, 12, 18, and 24. Given the six-fold symmetry of the ELT pupil and our choice of sampling, the other modes are all rotations of these unique modes.

At the higher separations, the modes begin to resemble that of the planet signal, with minor deviations to cancel out the star and also satisfy orthogonality with respect to the other modes. At closer-in separations, the modes become more distorted in order to maintain a low $\eta_s$. Although we have chosen samples separated by 1.22 $\lambda/D$ to perform measurements for, we note that the planet throughput at points in between the samples is mostly retained by the instrument, just split up between multiple modes. For example, a planet at $\xi_{p_x}=3.05 \lambda/D$ would fall exactly between two of the samples, but would have $>40\%$ throughput in each of Mode 6 and Mode 18 and about $10\%$ throughput in Mode 0, for a combined throughput of over $90\%$. For actual observations, the angle between the coronagraph and the sky can be modulated to fill in non-continuously sampled FOVs by allowing the field to rotate over a ground-based telescope, or by rolling a space telescope. This also provides angular diversity to calibrate potential quasi-static wavefront errors, though the data processing method has to be adapted to account for nonstandard, spatially-dependent PSFs (Goyal et al., under review).

\section{Discussion} \label{sec:discussion}

This work has focused on calculating optimal coronagraphs without considering the effect of wavefront error, but the density matrix can be modified to include potential wavefront errors by writing $\ket{\psi_s} = (1-\sqrt{\epsilon})\ket{\psi_{s_0}}+\sqrt{\epsilon}\ket{\delta \psi_s}$ and $\ket{\psi_p} = (1-\sqrt{\epsilon})\ket{\psi_{p_0}}+\sqrt{\epsilon}\ket{\delta \psi_p}$. When averaged over time, WFE averages out to two additional incoherent terms, one modifying the planet signal and one modifying the star. A mode-sorting coronagraph can be designed to compensate for the WFE, providing the optimal measurement given its statistics. However, the performance will be worse than if the wavefront error did not exist, as a coronagraph that suppresses the light caused by the wavefront error will also have reduced planet throughput. Instead of designing coronagraphs to accommodate all of the expected WFE, one can use wavefront sensors to measure the light not measured by the coronagraph, constraining the WFE associated with each frame of data and allowing for its effects to be removed in post-processing. Several implementations of this approach, referred to as wavefront calibration or PSF calibration, have been demonstrated using various instruments. However, if the WFE increases the average intensity measured, then it still results in additional photon noise. Finding the balance between maintaining planet throughput and reducing the photon noise associated with WFE --- as well as optimally decomposing the wavefront for PSF calibration --- is a topic for future work.

It is also possible to frame the coronagraphy problem as one that optimizes the QFI for planet position rather than the planet flux. However, this is more complicated, as the derivatives of $\ket{\psi_{pj}}$ with respect to the spatial parameters $\vec{\xi_p}$ will couple into other modes. Most use cases of coronagraphy tend to prioritize constraining the planet flux as well as possible, and optimally measuring the location of the planet is usually going to be incompatible with optimally measuring the planet flux. However, observational techniques such as spectroastrometry may benefit from a form of coronagraphy that optimizes the measurement of small astrometric deviations away from a point source rather than of the flux of the off-axis sources \citep{bailey_spectroastrometry_1998}. Mode-sorting spectroastrometry with photonic lanterns in particular has already been conceptualized and demonstrated on-sky \citep{kim_-sky_2025}, with the potential for nulling variants to improve the sensitivity limits of the technique.

Lastly, future work will include research on manufacturing devices that can sort the user-defined modes needed for optimal coronagraphy. We plan to explore the use of a multi-plane light converter \citep{morizur_programmable_2010, labroille_efficient_2014}, which propagates the light through a series of phase screens that converts it between different mode bases. The phase screens needed to sort the optimal modes can be calculated using inverse design methods, and the phase screens themselves can manufactured using, for example, direct-write liquid crystal technology \citep{doelman_patterned_2017}. Photonic lanterns or photonic chips of beam-combiners (or a combination of both) may also be able to realize quantum optimal coronagraphs. One could hypothetically approximate such a coronagraph using a pupil-reshaping optics \citep{guyon_phase-induced_2003} and a series of phase masks in front of fibers, where the light rejected by the first phase mask is sent to the second phase mask and so on, but this is possibly too complicated and unstable to be a realistic implementation. Ultimately, the coronagraphic performance of any of these devices will depend on how well manufacturing methods can produce the designs needed for sorting the optimal modes.

\section{Conclusions}

In this work, we show that concepts from quantum measurement provide powerful tools for characterizing the theoretical limits of coronagraphy and for calculating the optimal coronagraph for a given measurement task. Because optimal measurements of multiple planets are usually mutually incompatible, coronagraphs with a finite FOV will involve inherent tradeoffs between the planet locations targeted and the sensitivity that is achievable at those locations. Therefore, as we show through several example coronagraph designs, there is no single optimal coronagraph, and the desired properties of the coronagraph must be tuned for the particular science case of interest.

\begin{acknowledgements}
The authors acknowledge support from NWO Award 184.036.004.
\end{acknowledgements}

%

\bibliographystyle{bibtex/aa}
\bibliography{references}

\begin{appendix}





\onecolumn
\section{Primer on quantum sensing terminology} \label{app:qm_primer}
This section is a primer on quantum sensing terminology and notation intended for astronomers. We assume that the reader is familiar with undergraduate level quantum mechanics, including the description of quantum states using either wavefunctions or complex vectors in a Hilbert space. We begin by reviewing bra-ket notation, which is used extensively throughout this work.

\subsection{Quantum states as complex vectors}
A particle, such as a photon received by a telescope, exists in a quantum state, which can be described by a vector in Hilbert space. These states are denoted as a ket $\ket{\psi}$. In the field of coronagraphy, the state of the light is commonly simplified to a description of the scalar value of the electric field at a given plane along the axis of propagation through the instrument. At the entrance of the telescope pupil, this electric field state of a photon from the onsky position of $\vec{\xi}$ is given by the complex vector

\begin{equation}
    \ket{\psi(\vec{\xi})} = E(\vec{\xi}) = Ae^{i\vec{\xi} \cdot \vec{x}},
\end{equation}

where $A$ is the function describing the telescope aperture. The complex conjugate of the vector is denoted as a bra $\bra{\psi(\vec{\xi})} = E^{\dagger}(\vec{\xi})$. The complex inner product between two states can be denoted as a bra-ket, and also be described with an overlap integral:

\begin{equation}
    \braket{\psi(\vec{\xi}_1)|\psi(\vec{\xi}_2)} = E^{\dagger}(\vec{\xi_1})E(\vec{\xi}_2) = \iint_D E^*(\vec{\xi_1}) E(\vec{\xi_2}) d^2\xi.
\end{equation}

\subsection{Mixed states and density matrices}
While single vector describes the true quantum state of a given particle, additional formalism is needed to describe a particle whose state is unknown, but drawn from a distribution of possible states. Such a particle is said to be in a mixed state, which is denoted with a density matrix

\begin{equation}
    \hat{\rho} = \sum_j p_j\ket{\psi_j} \bra{\psi_j}.
\end{equation}

The $p_j$ encode the probability that the state of the particle is in $\ket{\psi_j}$. The trace of $\hat{\rho}$ must necessarily be one, as the total probability must add up to one. The density matrix encodes both the probability information as well as the coherence information of the states involved. For example, if photons are received from two equally bright point sources on the sky, the density matrix would be

\begin{equation} \label{eq:equal_bin}
    \hat{\rho}_{\mathrm{bin}} = 0.5 \ket{\psi_1}\bra{\psi_1} + 0.5 \ket{\psi_2}\bra{\psi_2}.
\end{equation}

This density matrix reflects both the fact that a received photon has a 50$\%$ chance of being from either source (due to their equal brightness), and the fact that the photons emitted by one source are incoherent with those from the other, and thus do not interfere.

\subsection{Quantum measurements}
Quantum measurement can be described using measurement operators $\hat{M}_k$. The probability of detecting a photon using the measurement labeled by $k$, given the state $\hat{\rho}$, is

\begin{equation}
    p_k = \mathrm{Tr}(\hat{\rho}\hat{M}_k).
\end{equation}

See \citet{tsang_quantum_2016} for an explanation of why quantum measurements of light through a telescope correspond to measurements of spatial modes of light. Intuitively, the quantum states $\ket{\psi}$ of the incoming photons are spatial modes of light which form a Hilbert space, so it follows that the measurements $\hat{M}_k$ are also of spatial modes of light within that Hilbert space, i.e. $\hat{M}_k = \ket{\phi_k}\bra{\phi_k}$ for the measurement of the photons in a single spatial mode, or $\hat{M}_K = \sum_ka_k\ket{\phi_k}\bra{\phi_k}$ for a weighted sum of the photons across a subspace of modes. Even conventional imaging, or coronagraphy with an image-forming backend, involves measurements of spatial modes of light. The modes sorted by imagers are the top-hat modes corresponding to pixels of the detector, with possible measurements involving weighted sums of the photons across multiple pixels (e.g. photometry using an aperture or a matched filter).

For additional intuition about this description of quantum measurement, we can analyze the equal-brightness binary state $\hat{\rho}_{\mathrm{bin}}$ defined in Eq. \ref{eq:equal_bin}. For a spatial mode $\phi(\vec{\xi})$ and source electric fields $\psi_1(\vec{\xi})$ and $\psi_2(\vec{\xi})$ defined in the spatial coordinates of the focal plane, the measurement $p_k = \mathrm{Tr}(\hat{\rho}\hat{M}_k)$ can be written

\begin{equation}
    p_{k_{\mathrm{bin}}}(\phi) = 0.5 \bigg|\iint_D \phi^*(\vec{\xi}) \psi_1(\vec{\xi}) d^2\xi \bigg|^2 + 0.5 \bigg|\iint_D\phi^*(\vec{\xi}) \psi_2(\vec{\xi}) d^2\xi  \bigg|^2,
\end{equation}

where $\iint_D$ indicates an integration over the entire focal plane. The probability of measuring a photon in mode $\phi$ is thus a weighted sum of the overlap of $\phi$ with the electric fields corresponding to each of the two sources. Such overlap integrals are commonly used to describe the coupling of light into fibers, and one can see why fiber-based nulling instruments \citep{haguenauer_deep_2006, ruane_efficient_2018} were some of the earliest and most natural implementations of non-imaging-based mode-sorting coronagraphy.

\subsection{(Quantum) Fisher information and parameter estimation}

Once the data from a hypothetical measurement $\hat{M}_k$ are obtained, we can use the data to infer the value of a parameter $\theta$, encoded within the density matrix $\hat{\rho}(\theta)$. From classical estimation theory, we have the Fisher information given by

\begin{equation}
    F_c(\theta) = \sum_k \frac{1}{p_k} (\frac{\partial p_k}{\partial \theta})^2.
\end{equation}

For multiparameter estimation over a vector of parameters $\theta_{\mu}$, the Fisher information becomes a matrix with parameters

\begin{equation}
    F_{c_{\mu \nu}} = \sum_k \frac{1}{p_k} \frac{\partial p_k}{\partial \theta_\mu}\frac{\partial p_k}{\partial \theta_\nu}.
\end{equation}

The Fisher information is calculated for the set of $p_k$ provided by a given instrument, and it encodes the sensitivity of the instrument to the parameter we wish to estimate. Quantum measurement theory allows us to calculate the ultimate sensitivity limit given all possible instruments, as well as what kind of measurement is needed to achieve that limit. The quantum Fisher information (QFI) given by $F_Q(\theta)$ is not a different kind of information than the classical Fisher information. It is rather the supremum over possible $F_c(\theta)$, a quantity intrinsic to the state $\hat{\rho}(\theta)$ itself, and achievable given the theoretically optimal measurement.

The theoretically optimal measurement is given by the operator called the symmetric logarithmic derivative (SLD), which itself only depends on the state $\hat{\rho}(\theta)$ and its derivative with respect to $\theta$. It is defined implicitly as

\begin{equation}
    \partial_\theta\hat{\rho}(\theta) = \frac{1}{2} (\hat{L}_\theta \hat{\rho}(\theta)+\hat{\rho}(\theta)\hat{L}_\theta),
\end{equation}

and can be calculated in the eigenbasis of $\hat{\rho}(\theta)$ (with eigenvalues $\zeta_m$ and eigenvectors $\ket{\zeta_m}$) using the formula

\begin{equation}
    \hat{L}_\theta = 2 \sum_{m,n} \frac{\braket{\zeta_n|\partial_\theta \hat{\rho}(\theta)|\zeta_m}}{\zeta_m+\zeta_n} \ket{\zeta_n}\bra{\zeta_m}.
\end{equation}

For a single parameter, the QFI can be calculated from the SLD as

\begin{equation}
    F_Q(\theta) = \mathrm{Tr}(\hat{\rho}\hat{L}_\theta^2).
\end{equation}

For multiparameter estimation, the elements of the QFI matrix $F_{Q_{\mu \nu}}$ are given by

\begin{equation}
    F_{Q_{\mu \nu}} = \mathrm{Tr}(\hat{\rho}\hat{L}_{\theta_\mu}\hat{L}_{\theta_\nu}).
\end{equation}

However, as explored more in Section \ref{sec:fov_tradeoffs}, it may not be possible to simultaneously achieve the theoretical limit for multiple parameters, as the optimal SLDs for each parameter may not commute with one another.

Ultimately, these theoretical results from quantum measurement theory do not offer a fundamentally new way of performing measurements compared to our classical descriptions of optical instruments. However, they do provide useful computational tools for calculating and characterizing the information limits for tasks such as planet flux estimation, for designing instruments that can reach these fundamental limits, and for exploring the tradeoffs associated with instruments that must balance sensitivity across multiple measurement tasks.

\section{Relationship between quantum and classical optimality for flux estimation} \label{app:quantum_to_classical}

In this section, we relate the concept of the QFI to the classical concept of S/N ratio. The square of the classical S/N ratio in the presence of photon noise is typically written as $(S/N)^2=\eta_p^2/(\eta_s+c\eta_p)$, where $\eta_p$ is the throughput of the planet, $\eta_s$ is the throughput of the star, and $c$ the flux ratio of the planet. The expression for the classical S/N ratio uses the planet intensity as the signal and is thus analogous to the quantum measurement task of estimating the parameter $c$. The classical Fisher information (CFI) for the parameter $c$ with a projective measurement is given by the sum over possible outcomes $k$, each with probability $p(\phi_k|c)$,

\begin{equation} \label{eq:cfi_general}
    F_c(c) = \sum_k \frac{1}{p(\phi_k|c)}(\frac{\partial p(\phi_k|c)}{\partial c})^2.
\end{equation}

For an unresolved star and a single $\ket{\psi_p}$, the measurement has to lie in the space spanned by their respective modes and can be written as that of a projection onto the mode $\phi$ given by $\ket{\phi}\bra{\phi}$ and the projection onto its orthogonal space given by $\mathds{1} - \ket{\phi}\bra{\phi}$. We make here an assumption to simplify the calculation of the probability of measuring the state $\phi$ (i.e. detecting a photon) given $c$, which is that the center of intensity of the star is known and that the instrument can be aligned to it. This results in the states $\ket{\psi_s}$ and $\ket{\psi_p}$ being independent of $c$. Otherwise, if the instrument is aligned to the center of intensity between the star and a potential planet, then $\ket{\psi_s}$ and $\ket{\psi_p}$ are dependent on $c$, resulting derivatives that couple to other states.

How accurate our assumption is depends on where and how bright the non-stellar sources are, as well as on the alignment procedure used. For example, if alignment is done by centroiding the light within some window, then a bright companion that falls within the centroiding window would shift the center of intensity away from the star. If the planet is much fainter than the star, or if the planet is far away and falls outside of the window used, then this shift would negligible, and we can assume that the instrument is well-centered on the star. In this scenario, $\ket{\psi_s}$ and $\ket{\psi_p}$ are independent of $c$, and the probability of measuring $\phi$ given $c$ is

\begin{equation} \label{eq:p_phi}
    p(\phi,c)=\braket{\phi|\rho|\phi} = (1-c)|\braket{\phi|\psi_s}|^2+c|\braket{\phi|\psi_p}|^2,
\end{equation}

while the probability of measuring the perpendicular space is $p({\perp},c) = 1-p(\phi,c)$, so $(\partial p(\perp,c)/\partial c)^2 = (\partial p(\phi,c)/\partial c)^2$. The throughputs of the planet and star in the mode $\phi$ are given by $\eta_p=|\braket{\phi|\psi_p}|^2$ and $\eta_s=|\braket{\phi|\psi_s}|^2$ respectively. After some simplification, the CFI of this measurement is

\begin{equation}
    F_c(c) = \frac{(\eta_p-\eta_s)^2}{[(1-c)\eta_s+c\eta_p]
    [1-((1-c)\eta_s+c\eta_p)]},
\end{equation}

and the optimal measurement is the one that maximizes this expression. Within the coronagraphy context, generally $\eta_s \ll \eta_p$ and $c\ll \eta_p$, while $c$ is approximately equal to $\eta_s$ for the most challenging targets and greater than $\eta_s$ otherwise. For $\eta_s\ll\eta_p$, the numerator is approximately $\eta_p^2$. Similarly, the denominator to leading order in $\eta_s$ and $c$ (dropping terms of order $c^2$, $\eta_s^2$, $c\eta_s$, and higher) becomes the familiar $\eta_s+c\eta_p$. Therefore, in typical coronagraphy contexts, the Fisher information is approximately the square of the classical S/N ratio

\begin{equation}
    F_c(c) \approx \frac{\eta_p^2}{\eta_s+c\eta_p}.
\end{equation}

Part of the difference between them is that the classical S/N expression uses for the signal $c$ as the flux ratio instead of as the relative flux contribution, when they are only approximately equal in the limit of small $c$. The other part of the difference arises from the fact that the classical S/N ratio assumes that the binomial distribution can be approximated as Poissonian, with variance equal to the number of photons detected. The quantum optimal measurement maximizes the ability to constrain $c$ from the actual distribution of measurements across $\phi$ and $\phi_{\perp}$, which are that of a Bernoulli process: if $n$ photons are measured, the probability of measuring $\phi$ $k$ times follows the binomial distribution $f_{\mathrm{binomial}}(k,n,p) = {n\choose k} p^k(1-p)^{(n-k)}$, where $p=p(\phi,c)= (1-c) \eta_s + c \eta_p$ and $p(\phi_{\perp},c) = 1-p(\phi,c)$. The expectation value of photons detected in $\phi$ is $np$ with variance $np(1-p)$. Meanwhile, the Poisson distribution $f_{\mathrm{poisson}}=(\lambda^ke^{-\lambda})/k!$ assumed in the classical expression for S/N is an approximation of the binomial distribution describing the limiting case of large $n$ and small $p$, whereby $\lambda=np$ and the expectation value and variance of photons detected in $\phi$ are both $\lambda$. It turns out that in the case of an unresolved star, the quantum optimal measurement is orthogonal to $\ket{\psi_s}$ regardless of $c$. As shown numerically in Section \ref{sec:unresolved_star}, the mode that optimizes $(S/N)^2=\eta_p^2/(\eta_s+c\eta_p)$ is approximately quantum optimal for small $c$ but deviates from it as $c$ grows larger.

For finite sized stars, the argument is more nuanced as the stellar modes span all the modes of the telescope, which is related to the known fact that coronagraphs that satisfy $\eta_s=0$ exactly with a finite star will also have zero planet throughput \citep{guyon_theoretical_2006, belikov_theoretical_2021}. The complication arises because $\hat{\rho}(c)$ is no longer (at most) rank 2, so the optimal measurement given by the SLD is no longer just a projection onto a vector $\phi$ and its orthogonal component.

Here it is useful to note that given an infinitesimal variation away from the state containing only the star, $\hat{\rho}_1 = \hat{\rho}_0 + \delta \hat{\rho}$, the measurement that maximizes the statistical distinguishability between $\hat{\rho}_1$ and $\hat{\rho}_0$ is a projective measurement onto the eigenvectors of the matrix $\hat{\rho_0}^{1/2}\hat{\rho_1}\hat{\rho_0}^{1/2}$ \citep{fuchs_mathematical_1995}. This measurement maximizes the sensitivity to a perturbative $c$ in the limit of $c\rightarrow 0$. For a single planet, the operator $\hat{\rho_0}^{1/2}\hat{\rho_1}\hat{\rho_0}^{1/2}$ has the eigenvectors of $\hat{\rho_0}^{1/2}\ket{\psi_p}\bra{\psi_p} \hat{\rho_0}^{1/2}$, which is rank 1. Therefore, the optimal measurement in the $c\rightarrow0$ limit is still a projective measurement onto a mode $\phi$ and its orthogonal component, and the calculation for the Fisher information is the same as before. The difference is that the $(1-c) \ket{\psi_0}\bra{\psi_0}$ in the density matrix is replaced with the incoherent sum of possible stellar states. For a discrete model of the star with $N_s$ samples over the stellar disk, the stellar contribution to the density matrix is $\frac{1-c}{N_s} \sum_i^{N_s} \ket{\psi_{si}}\bra{\psi_{si}}$, and the stellar throughput becomes $\eta_s = \frac{1}{N_s}\sum_i^{N_s} |\braket{\phi|\psi_{si}}|^2$. The single $\phi$ that maximizes the classical S/N ratio given a finite star is still approximately quantum optimal, to leading order in $c$ and $\eta_s$. However, it is less straightforward to explain the deviation away from quantum optimality given a finite $c$, as the SLD for finite $c$ and finite stellar size generally has multiple nonzero eigenvalues and thus involves the measurement of more than one mode. We quantify the difference numerically in Section \ref{sec:single_planet_sld}.

\section{Classical optimization of optimal measurement mode} \label{app:classical_opt}

\subsubsection{Unresolved star} \label{sec:unresolved_star}

To calculate the mode $\phi$ that optimizes the conventionally used expression for the S/N ratio, we denote the signals of the planet and star (both infinitesimal point sources) as their electric fields $a=E_p$ and $b=E_s$ respectively, where both vectors have encoded within them the pupil shape. We will also normalize them such that $b^Hb=1$ and $a^Ha=1$, and to get the planet intensity, we will multiply $a^Ha$ by the flux ratio denoted by $c$. For a fixed star mode $b$ and a single known planet mode $a$, the mode $\phi$ that maximizes the square of S/N is given by

\begin{equation}
    \max_\phi \frac{(\phi^H a a^H \phi)^2}{(\phi^H b b^H \phi)+c(\phi^H a a^H \phi)}.
\end{equation}

The solution should be a linear combination of $a$ and $b$ and is thus of the form $\phi = \alpha a+\beta b$. It will also depend on the inner product of $a$ and $b$, which can be denoted as $\kappa = a^Hb $ (which implies $\bar{\kappa}=b^Ha$). Additionally, because of the constraint that $\phi^H\phi=1$, it is useful to parametrize the solution in terms of a complex ratio of $\alpha$ and $\beta$ given by $r = \beta/\alpha$. We now have various quantities in these parameters given by $\phi = \alpha(a+rb)$, $a^H\phi = \alpha(1+r\kappa)$, and $b^H\phi = \alpha(\bar{\kappa}+r)$. Plugging these values into the cost function gives

\begin{equation}
    f=|\alpha|^2\frac{(1+\kappa r)^2(1+\bar{\kappa}\bar{r})^2}{(\kappa+r)(\kappa+\bar{r})+c(1+\kappa r)(1+\bar{\kappa} \bar{r})}.
\end{equation}

The $|\alpha|^2$ in the cost function can be replaced using the constraint $\phi^H\phi=1$. Also using $a^Ha=1$ and $b^Hb=1$, this becomes $|\alpha|^2+|\beta|^2+\bar{\alpha}\beta a^Hb + \alpha \bar{\beta}b^Ha=1$. Plugging in $\beta = \alpha r$ gives $|\alpha|^2+|\alpha|^2|r|^2+|\alpha|^2r\kappa+|\alpha|^2\bar{r}\bar{\kappa} = 1$, or $|\alpha|^2 = 1/(1+r^2+r\kappa+\bar{r}\bar{\kappa})$.

The cost function in terms of just $r$ is now

\begin{equation}
    f=\frac{1}{(1+r^2+r\kappa+\bar{r}\bar{\kappa})}\frac{(1+\kappa r)^2(1+\bar{\kappa}\bar{r})^2}{((\kappa+r)(\bar{\kappa}+\bar{r})+c(1+\kappa r)(1+\bar{\kappa} \bar{r}))}.
\end{equation}

For a symmetric pupil, $\kappa$ is real ($\kappa=\bar{\kappa})$. We can also try to look for a solution where $r$ is also real ($r=\bar{r}$), using the cost function

\begin{equation}
    f=\frac{1}{(1+r^2+2r\kappa)}\frac{(1+\kappa r)^4}{((\kappa+r)^2+c(1+\kappa r)^2)}.
\end{equation}

Taking the derivative and setting the numerator to zero gives $2(\kappa-1)(\kappa+1)(r\kappa+1)^2(cr^3\kappa^2 + 2cr^2\kappa + cr + 2r^3+5r^2\kappa+3r\kappa^2 +r+\kappa) = 0$. The first two terms are independent of $r$, the second term results in $f=0$ being a minimum, so we would like $r_{\mathrm{opt}}$ to be the root of the third term, the polynomial equation

\begin{equation} \label{eq:unresolved_r_polynomial}
cr^3\kappa^2 + 2cr^2\kappa + cr + 2r^3+5r^2\kappa+3r\kappa^2 +r+\kappa = 0.
\end{equation}

Note that we have used the Poisson approximation here for simplicity. The steps of the derivation can be performed with the binomial distribution if desired, and would result in a different polynomial in Eq. \ref{eq:unresolved_r_polynomial}. 

\begin{figure*}[t]
\begin{center}
	a\includegraphics[scale = 0.48]{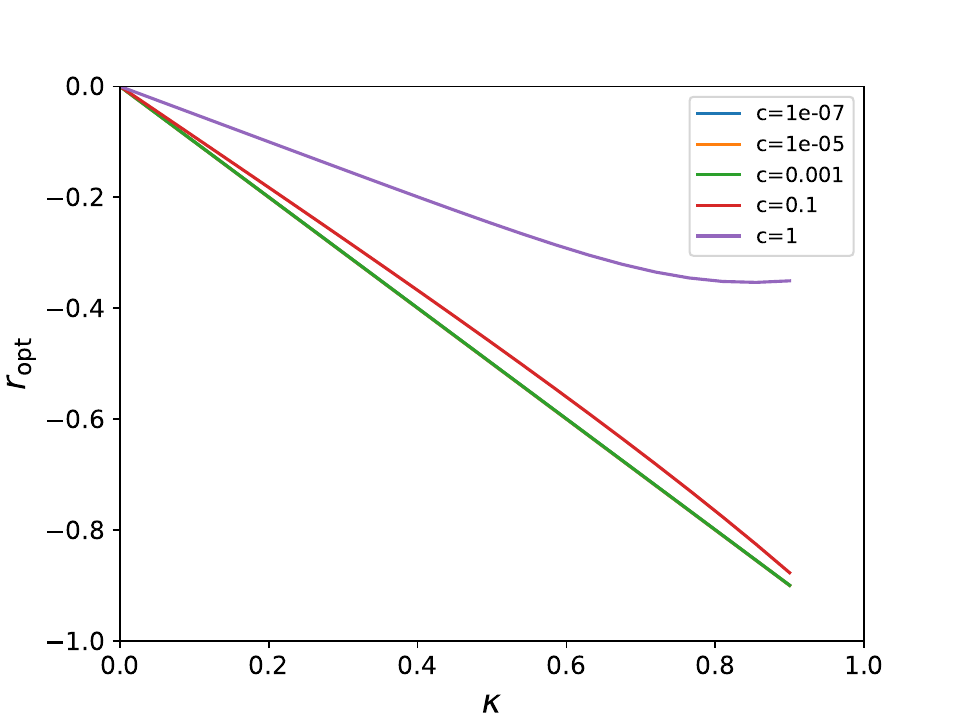}
    b\includegraphics[scale = 0.48]{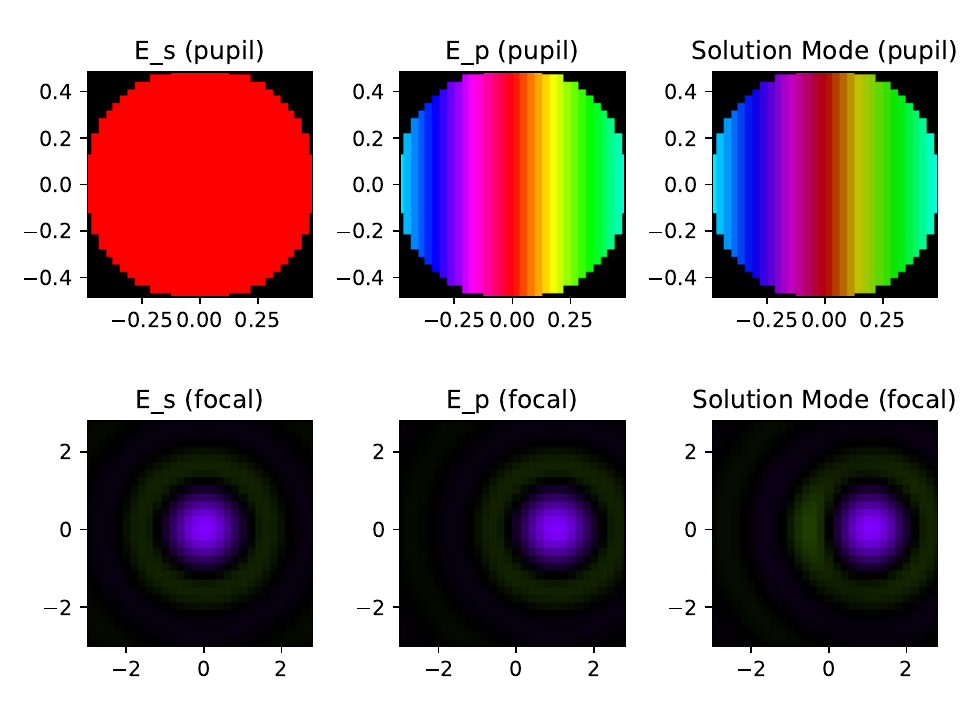}
	\caption{\label{fig:projection_solution} a) The classically-optimal ratio between stellar and planet modes ($r_{\mathrm{opt}}$) from Eq. \ref{eq:unresolved_r_polynomial}, as a function of $\kappa$, the overlap between the modes. For $c\ll1$, $r_{\mathrm{opt}}\approx\kappa$, which is the projection of the electric field along the component of the planet signal that is perpendicular to the stellar signal. b) An example solution for a planet with flux ratio $c=1\times 10^{-7}$ at a spatial position of $\xi_x=1$ and $\xi_y=0$. The mode $E_s$, the mode $E_p$, and the optimal mode-sorter mode $\phi_{\mathrm{opt}}$ that maximizes the S/N ratio are shown, in both the pupil and focal plane.}
\end{center}
\end{figure*}

The numerical real root $r_{\mathrm{opt}}$ of Eq. \ref{eq:unresolved_r_polynomial} exists for $\kappa$ smaller than $\sim 0.9$. This is fortunately the range of $\kappa$ that we are practically interested in, since even a planet as close as $0.5 \lambda/D$ has only $\kappa\approx0.72$ given an unobstructed circular pupil. We see that the classical solution agrees with the quantum optimal solution (a projection $\ket{\psi_s}$ and the component of $\ket{\psi_p}$ orthogonal to $\ket{\psi_s}$) even up to values of $c$ as large as $\sim0.1$. The value of $r_{\mathrm{opt}}$ given an unobstructed circular pupil for a range of $c$ are plotted in Figure \ref{fig:projection_solution}a. The ratio $r_{\mathrm{opt}}$ can be converted back into the coefficients $\alpha$ and $\beta$ to find the actual desired mode. The solution for a planet with flux ratio $c=10^{-7}$ at a spatial position of $\xi_x=1$ and $\xi_y=0$ is shown in Figure \ref{fig:projection_solution}b.

\subsection{Resolved star}

For finite sized stars, again using the Poisson approximation, the cost function for maximizing classical S/N in a mode $\phi$ is

\begin{equation} \label{eq:classical_L_finite}
    \max_\phi \frac{(\phi^H a a^H \phi)^2}{\int\int(\phi^H b(\vec{\xi}) b^H(\vec{\xi}) \phi)d\xi_x d\xi_y+c(\phi^Haa^H\phi)},
\end{equation}

where the integral is over the area extended by the star with radius $R_s$. With numerical calculations in mind (where the integral becomes a sum over discrete points index by $i$), the star's intensity is normalized such that $\int\int b^H(\vec{\xi})b(\vec{\xi})d\xi_xd\xi_y=\sum_i b_i^H b_i =1$. Adding the constraint of $\phi^H\phi=1$ through a Lagrange multiplier, we have:

\begin{equation} 
    L=\frac{(\phi^H aa^H\phi)^2}{(\sum_i\phi^Hb_ib_i^H\phi)+c\phi^Haa^H\phi}+\lambda(1-\phi^H\phi)
\end{equation}

Denoting $A=aa^H$, $B=\sum_i b_ib_i^H$, $G = \phi^H aa^H\phi$, and $H=(\sum_i\phi^Hb_ib_i^H\phi)+c\phi^Haa^H\phi$, and setting $\nabla L =0$ , we have $[2(G/H)A-(G/H)^2(B+cA)]\phi = \lambda \phi$. We can then absorb one factor of $G/H$ into the right-hand side with $\tilde{\lambda} = \lambda/(G/H)$ to get

\begin{equation}
    [2A-(G/H)(B+cA)]\phi = \tilde{\lambda} \phi.
\end{equation}

This is a nonlinear equation, as both $G$ and $H$ depend on $\phi$, and will require numerical methods to solve. Writing $\mu=G/H$ lets us solve a separate eigenvalue problem for all possible values of $\mu$, then we find the value of $\mu$ for which the quantity $G/H$ (computed using the relevant eigenvector) is actually equal to $\mu$. The eigenvector at this value of $\mu$ is the $\phi$ that optimizes Eq. \ref{eq:classical_L_finite}.

\begin{figure}[ht!]
\begin{center}
    Classically optimized solutions for finite star ($R_s=0.2$)\\ Planet at $\xi_{x_p}=1.0$\\
    \vspace{0.5em}
    \includegraphics[scale=0.43]{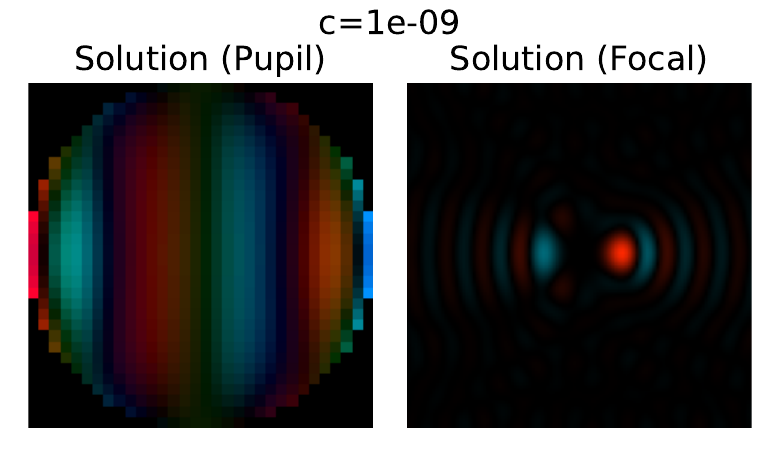} \hspace{1em}
    \includegraphics[scale=0.43]{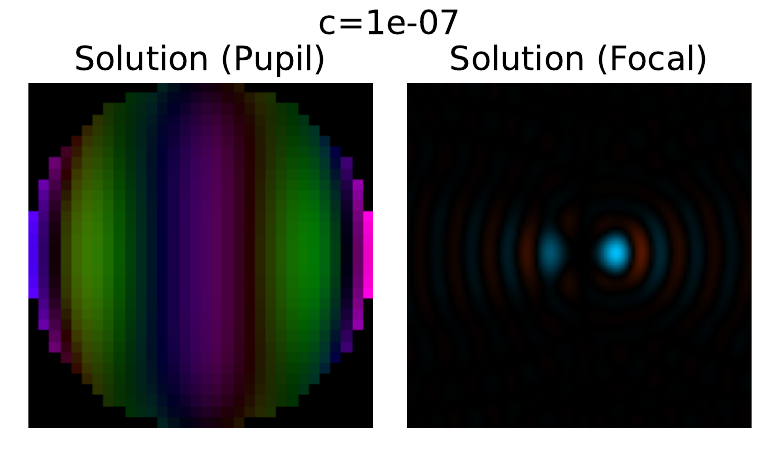} \hspace{1em}
    \includegraphics[scale=0.43]{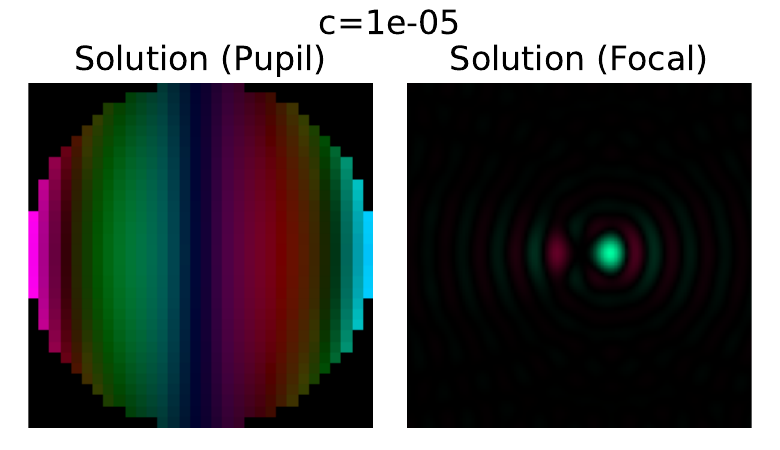}
	
	\caption{\label{fig:finite_star_sols} Examples of classically optimized mode solutions for $R_s=0.2$ and $\xi_{x_p}=1.0$ for flux ratios of $c=10^{-9}$, $c=10^{-7}$, and $c=10^{-5}$. For faint planets, it is more optimal to null the star at the expense of planet throughput. As the planet becomes brighter, retaining the planet light becomes more optimal.}
\end{center}
\end{figure}

We use numerically calculate this classically-optimized solution for the same astrophysical system as in in Sec. \ref{sec:single_planet_sld}, in which the star has radius $R_s=0.2$ and the planet is at a location of $\xi_{x_p} = 1.0$, also with a circular pupil. The solutions for a range of $c$ are shown in Fig. \ref{fig:finite_star_sols}. As discussed in App. \ref{app:numerical_disc}, calculating the quantum-optimal SLD is numerically difficult when it becomes effectively singular. In this regime, the classically-optimized solution shown here remains a viable calculation, and provides a very good approximation of the quantum optimal measurement.

\FloatBarrier
\section{Modes of the PCS PGM Survey Coronagraph} \label{app:pcs_pgm_modes}

\begin{figure*}[t]
\begin{center}
    \includegraphics[scale=0.39]{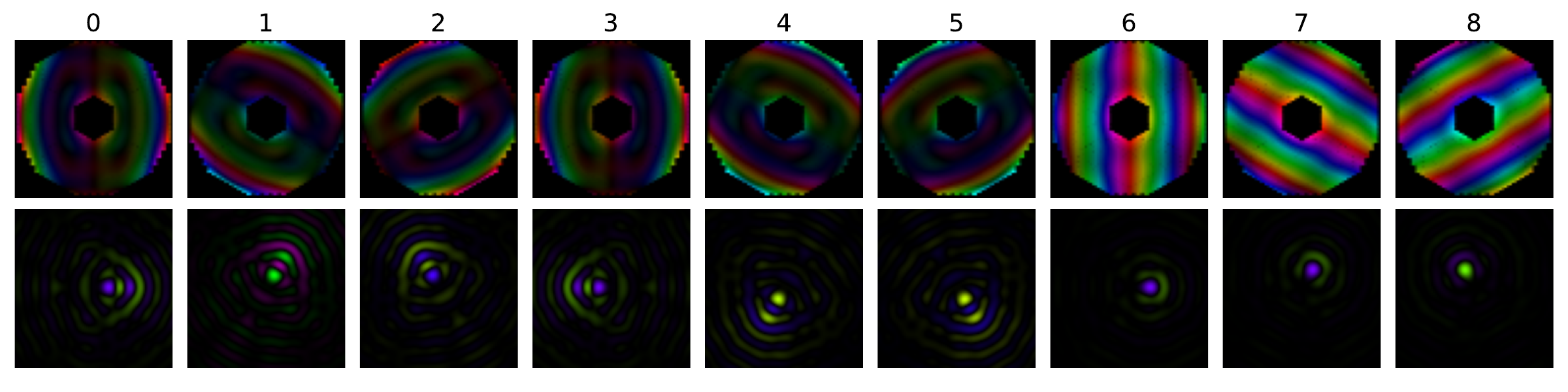}\\
    \includegraphics[scale=0.39]{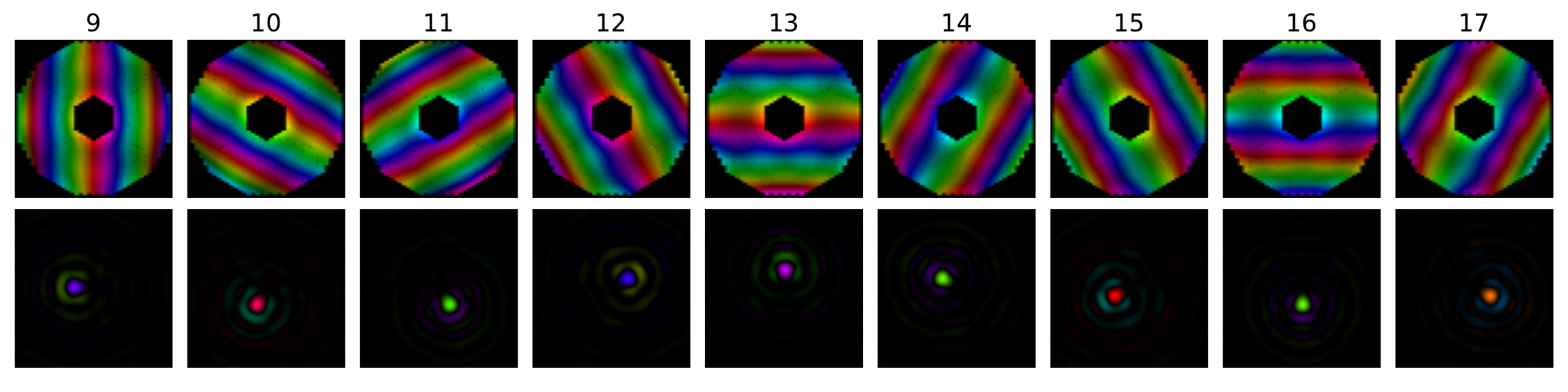}\\
    \includegraphics[scale=0.39]{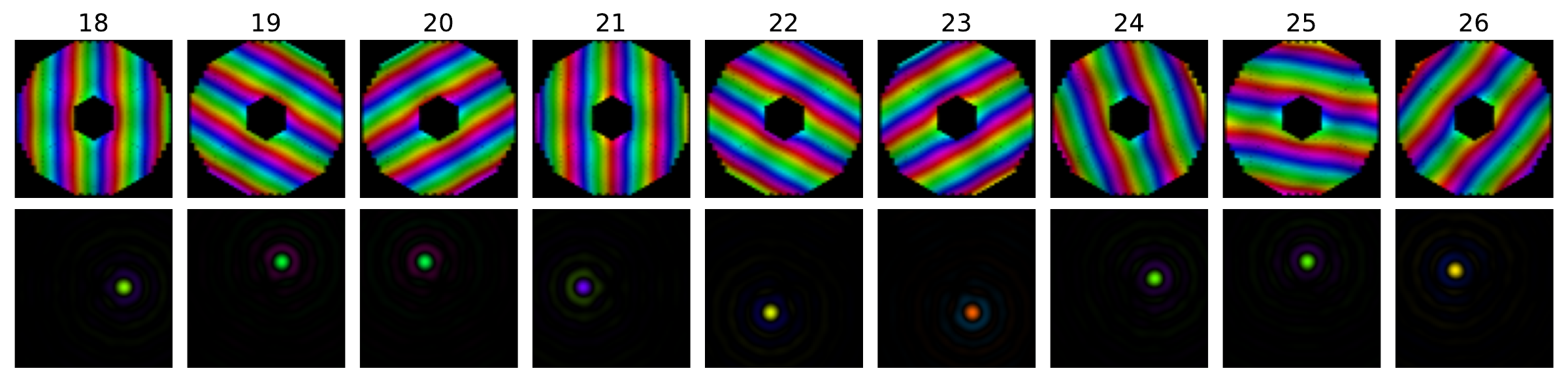}\\
    \includegraphics[scale=0.39]{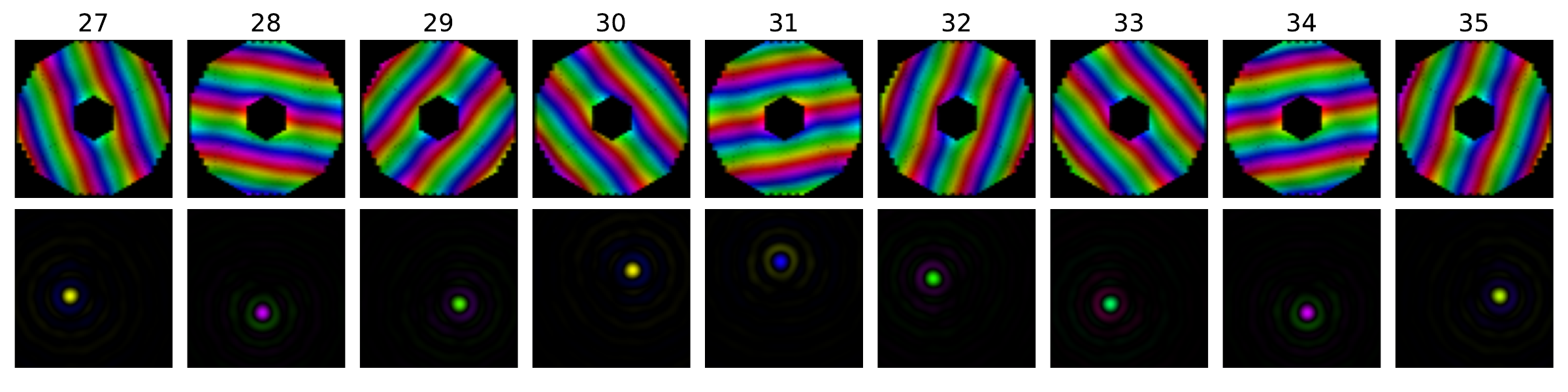}\\
	\caption{\label{fig:pcs_surveyor_modes} The PGM modes for the FOV samples in Fig. \ref{fig:pcs_surveyor_fov} given the ELT aperture, a star with radius $R_s=0.1 \lambda/D$, and $c=10^{-8}$. The pupil modes (top panel for each mode) are plotted from -0.5 to 0.5. The focal plane modes (bottom panel for each mode) are plotted from -10 to 10 $\lambda/D$. Given the six-fold symmetry of the ELT pupil and our FOV samples, there exist only five unique modes that can be rotated to attain the others. We present the modes grouped by symmetry, i.e. modes 1-5 are rotations of mode 0, modes 7-11 rotations of mode 6, modes 13-17 rotations of mode 12, modes 19-23 rotations of mode 18, and modes 25-35 rotations of mode 24. The throughput cross-sections along the axis intersecting the relevant FOV sample are shown for the unique modes in Fig. \ref{fig:pcs_surveyor_xsec}.}
\end{center}
\end{figure*}

\FloatBarrier

\section{Additional Notes}

\subsection{Coronagraphs for optimal detection} \label{app:chernoff_deriv}
Given the state of the star $\hat{\rho_0}$ and the mixed state $\hat{\rho_1} = (1-c)\hat{\rho}_0+c\ket{\psi_p}\bra{\psi_p}$, we could for example, calculate the measurement that, when repeated, achieves the maximum classical Chernoff exponent defined in Eq. \ref{eq:classical_chernoff}. For demonstrative purposes, we will assume a two-outcome set of measurements $E_\phi = \ket{\phi}\bra{\phi}$ and $E_\perp=1-\ket{\phi}\bra{\phi}$. We have that $p_0 =p_0(\phi)=\braket{\phi|\hat{\rho}_0|\phi} = \eta_s$, $p_1 =p_1(\phi)=\braket{\phi|\hat{\rho_1}|\phi} = (1-c)\eta_s+c\eta_p$, $p_0(\perp)= 1-p_0(\phi)$, and $p_1(\perp) = 1-p_1(\phi)$. Plugging these expressions into Eq. \ref{eq:classical_chernoff} gives

\begin{equation}
    \xi_{\mathrm{CB}} = -\log [\min_{0\leq s \leq 1}(1-p_0)^s(1-p_1)^{1-s}+p_0^s p_1^{1-s}].
\end{equation}

To calculate the minimum over $s$, we can take the derivative of the expression inside the log, set it to zero, and solve for $s^*$. The closed form solution for $s^*$ is complicated and involves logarithms of $p_0$ and $p_1$, but it exists so that the Chernoff exponent can be written as $\xi_{\mathrm{CB}}= -\log \chi(\phi)$ where $\chi(\phi) = (1-p_0(\phi))^{{s^*}(\phi)}(1-p_1(\phi))^{1-{s^*}(\phi)}+p_0(\phi)^{s^*(\phi)} p_1(\phi)^{1-{s^*}(\phi)})$. To maximize $\xi_{\mathrm{CB}}$, one must take its derivative and solve for the $\phi$ that satisfies $\partial_\phi\xi_{\mathrm{CB}} = \chi(\phi) \partial_\phi \chi(\phi)=0$. We can see that this is very complicated, though it is possible in principle. In practice, the maximum information measurement is much friendlier to calculate and interpret. We note that even when the classical Chernoff exponent is not used directly as the cost function for optimizing $\phi$, one can still plug in various potential $\phi$ (calculated using other methods) to compare their performance for hypothesis testing, if desired. For practical coronagraph design, however, where the data is typically not processed in a way that achieves the optimal classical Chernoff exponent, it is much simpler to calculate the measurement that is optimal for estimating the parameter $c$.

\subsection{QFI for various telescope apertures} \label{app:apertures}

\begin{figure*}[t]
\begin{center}
    \includegraphics[scale=0.55]{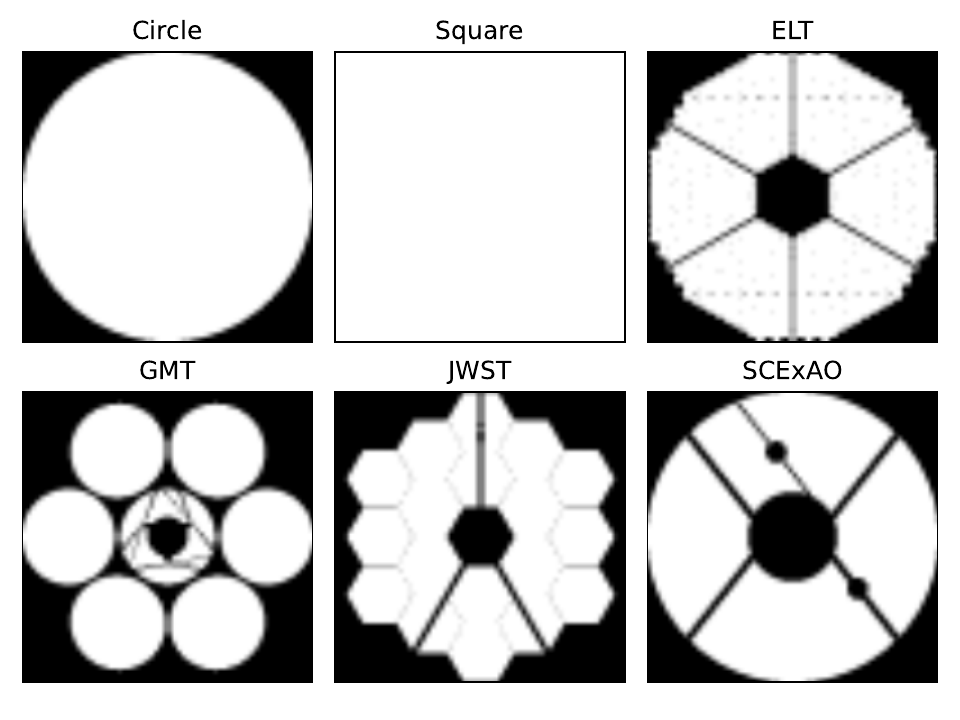}
    \includegraphics[scale=0.55]{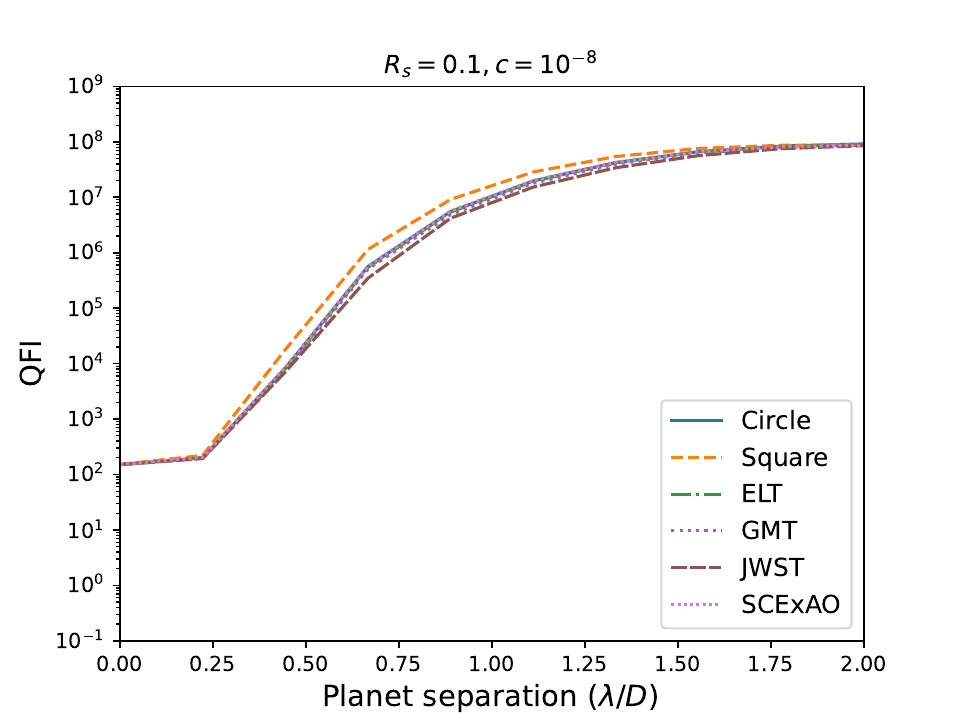}
	\caption{\label{fig:qfi_for_apertures} The QFI as a function of separation (on right) for various telescope apertures (on left, plotted from -0.5 to 0.5 on both axes). The stellar radius is assumed to be $R_s=0.1$, and the flux ratio $c=10^{-8}$. As predicted by previous theoretical studies \citep{guyon_theoretical_2006,belikov_theoretical_2021}, the aperture shape --- including the presence of spiders or obscurations --- has minimal impact on the theoretical limits of coronagraphy.}
\end{center}
\end{figure*}

In Fig. \ref{fig:qfi_for_apertures}, we compare the QFI of various telescope apertures for measuring the planet flux ratio given a particular star-planet system. Our calculations show that for a range of realistic shapes, the aperture function has only a minimal impact on the theoretical sensitivity limits. Only the square aperture shows a marginally higher sensitivity than the other aperture shapes, but that is due to the factor of $\sqrt{2}$ in spatial extent gained from spanning the entire pupil plane grid. Scaling the square aperture diameter down by $\sqrt{2}$ reduces its QFI curve to one that is marginally lower than the rest.

\subsection{Fragility of numerical calculations} \label{app:numerical_disc}
We have already discussed in Section \ref{sec:fov_tradeoffs} the numerical issues encountered while calculating the quantumness parameter $R$ describing measurement incompatibility. We would also like to note that the calculation of the SLD itself is also numerically fragile beyond planet separations of 2 to 3 $\lambda/D$, depending on the value of the stellar radius and flux ratio. We have limited our SLD calculations in this work to separations within $2 \lambda/D$, where the calculation is robust for all of the stellar radii and flux ratios we have explored. Fortunately, this encapsulates the range of astrophysical systems that conventional coronagraphs still struggle to access, which are of particular interest to the exoplanet community in the HWO and ELT era. However, directly applying the same computation for systems with very small stars ($R_s\ll 0.01)$, or very low flux ratios $c\ll10^{-10}$ --- especially with planet separations greater than $2 \lambda/D$ --- are likely to result in a calculation contaminated by numerical noise.

These numerical issues are not a consequence of our specific approach, but of the underlying physics of the problem. In exoplanet imaging, the extent of the star is quite small relative to the planet-star distance, but a star of any finite size is still physically full-rank on the space of the telescope. The faintness of the planet means that its presence causes only a tiny deviation in the density matrix describing the state of the overall system. This results in a situation where the numerical matrix describing $\hat{\rho}$ is extremely ill-conditioned. Yet, we cannot regularize away the modes with small singular values, as it is those modes that contain valuable planet signal. And, the further the planet is from the star, the higher order the modes the planet signal will populate. These combined factors result in SLDs that are nearly singular, but whose modes depend greatly on those high order modes of $\hat{\rho}$ with very low singular values. Increasing the precision of the calculations from that of the complex128 data type, by using the mpmath package for example, may increase the range of systems for which the SLD can be semi-analytically calculated. However, from a practical standpoint, it is usually not necessary to do so. As shown numerically in Section \ref{sec:single_planet_sld}, in the regime where the semi-analytic SLD calculation breaks down, the quantum optimal measurement is already effectively one-moded, and the shape of the mode can be calculated using conventional optimization techniques described in Section \ref{app:classical_opt}.

\FloatBarrier 
\clearpage

\end{appendix}
\end{document}